\documentclass[%
reprint,
superscriptaddress,
nofootinbib,
 amsmath,amssymb,
 aps,
pra,
]{revtex4-2}

\usepackage{graphicx}
\usepackage{dcolumn}
\usepackage{bm}
\usepackage{braket}
\usepackage{upgreek}
\usepackage{xcolor}
\usepackage{afterpage}
\usepackage{relsize}
\usepackage{multirow}
\usepackage{float}
\usepackage{physics}
\usepackage{makecell}
\usepackage{placeins}
\usepackage{amsthm}
\usepackage[colorlinks=true,urlcolor=blueprl,citecolor=blueprl,linkcolor=blueprl]{hyperref}
\usepackage{lineno}
\usepackage[symbol]{footmisc}
\usepackage{hhline}
\usepackage[normalem]{ulem}

\newcommand{\ozlem}[1]{\textcolor{black}{#1}}

\newcommand{\ozlemREV}[1]{\textcolor{black}{#1}}

\newcolumntype{P}[1]{>{\centering\arraybackslash}p{#1}}

\usepackage{hyperref}
\let\svthefootnote\thefootnote
\newcommand\freefootnote[1]{%
  \let\thefootnote\relax%
  \footnotetext{#1}%
  \let\thefootnote\svthefootnote%
}

\definecolor{blueprl}{RGB}{46,48,146}

\def\UQA{Centre of Excellence for Quantum Computation and Communication Technology, School of Mathematics and Physics, University of Queensland, St Lucia, QLD 4072, Australia}
\def\ASTAR{A*STAR Quantum Innovation Centre (Q.InC), Agency~for~Science,~Technology~and~Research~(A*STAR), 2 Fusionopolis Way, Innovis \#08-03, Singapore 138634, Republic of Singapore}
\def\ANU{Centre of Excellence for Quantum Computation and Communication Technology, The~Department~of~Quantum Science and Technology, Research School of Physics and Engineering, The Australian National University, Canberra, Australian Capital Territory, Australia}
\def\QCTRL{Present address: Q-CTRL, Sydney, New South Wales, Australia}

\begin{document}


\title{Utility of noiseless linear amplification and attenuation in single-rail discrete-variable quantum communications}
\author{\"{O}zlem Erk{\i}l{\i}\c{c}$^{*}$}
\freefootnote{$^*$ \href{ozlemerkilic1995@gmail.com}{ozlemerkilic1995@gmail.com}}
\affiliation{\UQA}

\author{Aritra Das}
\affiliation{\ANU}

\author{Angela A. Baiju}
\affiliation{\ASTAR}

\author{Nicholas Zaunders}
\affiliation{\UQA}

\author{Biveen Shajilal}
\affiliation{\ASTAR}
\affiliation{\QCTRL}


\author{Timothy C. Ralph}
\affiliation{\UQA}

\date{\today}
             

\begin{abstract}
    Quantum communication offers many applications, with teleportation and superdense coding being two of the most fundamental. In these protocols, pre-shared entanglement enables either the faithful transfer of quantum states or the transmission of more information than is possible classically. However, channel losses degrade the shared states, reducing teleportation fidelity and the information advantage in superdense coding. Here, we investigate how to mitigate these effects by optimising the measurements applied by the communicating parties. We formulate the problem as an optimisation over general positive operator-valued measurements~(POVMs) and compare the results with physically realisable noiseless attenuation~(NA) and noiseless linear amplification~(NLA) circuits. For teleportation, NLA/NA and optimised POVMs improve the average fidelity by up to $78\%$ while maintaining feasible success probabilities. For superdense coding, they enhance the quantum advantage over the classical channel capacity by more than $100\%$ in some regimes and shift the break-even point, thereby extending the tolerable range of losses. Notably, the optimal POVMs effectively reduce to NA or NLA, showing that simple, experimentally accessible operations already capture the essential performance gains.
\end{abstract} 

\maketitle

\section{\label{sec:introduction}Introduction}
Quantum communications harness the principles of quantum mechanics to enable information transfer with capabilities beyond those of classical channels. Over the past decades, this field has matured into a central pillar of quantum information, with applications ranging from quantum key distribution~(QKD)~\cite{ch1984quantum, ekert1991quantum, ralph1999, hillery2000, lucamarini2018, pirandola2013cvmdi, erkilicc2023surpassing, erkilic2025enhanced,erkilic2025software} for secure communications, to quantum teleportation~\cite{bennett1993teleporting, bouwmeester1997experimental, landry2007quantum, ma2012experimental, valivarthi2016quantum, valivarthi2020teleportation, braunstein1998teleportation, furusawa1998unconditional,  bowen2003experimental, takei2005high, yukawa2008high, lee2011teleportation, wang2021high, zhao2023enhancing} for state transfer, and superdense coding~\cite{barenco1995dense, mattle1996dense, barreiro2008beating, williams2017superdense, hu2018beating, braunstein2000dense, ralph2002unconditional, samanta2024continuous} for enhanced channel capacity.
\ozlem{Quantum communication protocols can be realised using two broad optical platforms: discrete-variable~(DV) encodings~\cite{gisin2002quantum} and continuous-variable~(CV) encodings~\cite{weedbrook2012gaussian}. In DV systems, information is stored in finite-dimensional photonic degrees of freedom, which can be encoded in two main ways. In single-rail DV~(SR-DV) encodings~\cite{lund2002nondeterministic, kok2007linear, kim2014transfer, jeong2016quantum, caspar2020heralded, polacchi2024teleportation}, the qubit is carried by a single optical mode, where the logical state correspond to the absence or presence of a photon. In contrast, dual-rail DV~(DR-DV) encodings~\cite{kok2007linear, bouwmeester1997experimental, mattle1996dense, barenco1995dense} distribute the qubit across two orthogonal modes such as two spatial paths or polarisation modes, so that the information is determined by which mode contains the photon. By contrast, CV platforms encode information in the continuous amplitude and phase quadratures of the optical field~\cite{weedbrook2012gaussian}.} 

While both DV and CV platforms provide powerful routes to implementing quantum communications, their performance is fundamentally constrained by transmission losses and channel noise. \ozlem{One route to overcoming these limitations is the use of quantum repeaters, which make more effective use of the distributed entanglement by combining entanglement swapping, heralding, and purification to extend communication distances despite channel loss~\cite{briegel1998quantum, dur1999quantum, munro2015inside, azuma2022quantum}. \ozlem{Major hurdles remain to implementing a scalable repeater system \cite{wang2025scalable}}. Another approach is to distribute entanglement through free-space links or satellite-based channels, which can outperform fibre distribution for long-distance communications~\cite{ursin2007entanglement, yin2017satellite1, yin2017satellite2, slussarenko2022quantum}. In realistic satellite channels, where Alice and Bob are ground stations and Charlie is the satellite node, a recent work~\cite{zaunders2025entanglement} has shown that directly distributing SR-DV entanglement from the satellite to both users outperforms relay-style configurations in which an entangled state is first prepared on the ground and routed through the satellite.}

\ozlem{The entangled states distributed through these channels form the resource for foundational quantum communication protocols such as teleportation and superdense coding.} Quantum teleportation enables the transfer of unknown quantum states between distant parties~\cite{bennett1993teleporting, bouwmeester1997experimental, landry2007quantum, ma2012experimental, valivarthi2016quantum, valivarthi2020teleportation, braunstein1998teleportation, furusawa1998unconditional, bowen2003experimental, takei2005high, yukawa2008high, lee2011teleportation, wang2021high, zhao2023enhancing}, while superdense coding allows the transmission of two classical bits using a single qubit in the presence of shared entanglement in the DV setting~\cite{barenco1995dense, mattle1996dense, barreiro2008beating, williams2017superdense, hu2018beating}. By contrast, CV encodings can in principle exceed two bits, with the achievable rate depending on the mean photon number~\cite{braunstein2000dense, ralph2002unconditional, samanta2024continuous}. In DV schemes, the two communicating parties typically pre-share a Bell state. For teleportation, Alice interferes the unknown state to be transmitted with her half of the entangled pair, performs a Bell-state measurement, and sends the outcome to Bob, who applies the appropriate correction to his qubit~\cite{bennett1993teleporting, bouwmeester1997experimental, landry2007quantum, ma2012experimental, valivarthi2016quantum, valivarthi2020teleportation}. In DV superdense coding~\cite{barenco1995dense, mattle1996dense, barreiro2008beating, williams2017superdense, hu2018beating}, Alice applies local operations to her qubit of the pre-shared entangled state to encode classical information and then sends it to Bob, who decodes the two bits via a Bell-state measurement. However, in both DV teleportation and superdense coding, losses or noise in the pre-shared entangled state reduce performance, leading to teleportation fidelities below unity and superdense coding capacities below the ideal two bits. One approach to improving the quality of pre-shared entanglement is entanglement purification. In this process~\cite{bennett1996purification, bennett1996mixed, deutsch1996quantum, horodecki1997inseparable, sheng2010deterministic, sheng2010one, sheng2013hybrid, hu2021long, winnel2022achieving, huang2022experimental, erkilic2024capacity}, multiple noisy entangled pairs are locally processed and measured by the communicating parties to distill a smaller set of states with higher purity. However, entanglement purification is highly resource-intensive and often requires multiple rounds of operation to achieve high fidelities.

On the other hand, in CV teleportation, the shared resource between the parties is a two-mode squeezed vacuum~(TMSV) state~\cite{weedbrook2012gaussian}. Alice mixes her half of the TMSV with the input state and performs a dual-homodyne measurement, whose outcomes are sent to Bob for a corresponding phase-space displacement~\cite{braunstein1998teleportation, furusawa1998unconditional, bowen2003experimental, takei2005high, yukawa2008high, lee2011teleportation, wang2021high, zhao2023enhancing}. Perfect fidelity would require infinitely squeezed TMSV states and lossless channels, which is unattainable in practice. To mitigate this, a recent CV teleportation experiment achieved a fidelity of $92\%$ by emulating noiseless linear amplification (NLA) through post-selection on dual-homodyne outcomes~\cite{zhao2023enhancing}. Physical NLAs, typically realised using quantum-scissor modules with single-photon detectors, ancilla photons, and beamsplitters~\cite{ralph2008nondeterministic, xiang2010heralded, ferreyrol2010implementation, kocsis2013heralded, winnel2020generalized}. \ozlemREV{They can also be implemented via conditional photon addition and subtraction, which provides an alternative experimentally demonstrated route ~\cite{zavatta2011high, neset2025experimental}. NLA-based schemes} have been shown to enhance the performance of both CV and DV systems~\cite{xiang2010heralded, blandino2012improving, ghalaii2020long, winnel2022achieving, notarnicola2023long, zaunders2025entanglement}. Recent concurrent work~\cite{jeong2026entanglement} has also studied heralded recovery of entanglement degraded by amplitude damping and photon loss in repeater architectures using weak-measurement reversal operations, noting that in photonic settings such reversal can be implemented via NLAs. Attenuation-type conditional operations have likewise been investigated experimentally using heralding on zero photons, also known as zero-photon subtraction~\cite{nunn2021heralding, nunn2022modifying}, with a range of applications in quantum information processing. In CV settings, NLA and noiseless attenuation (NA) are often implemented virtually via post-selection on measurement outcomes~\cite{fiuravsek2012gaussian, walk2013security, haw2016surpassing, zhao2017characterization, zhao2020high, hosseinidehaj2020finite, shajilal2024improving, erkilic2025enhanced, erkilic2025software}. However, such post-selection techniques cannot be directly applied in DV teleportation or superdense coding, where information is extracted via single-photon detection rather than continuous quadrature measurements.

\ozlem{In this work, we focus on the SR-DV setting and investigate how distributed entanglement can be used more effectively in the presence of lossy channels for both teleportation and superdense coding.} We formulate the task as an optimisation problem, maximising the output fidelity over a given set of input states in \ozlem{SR-DV} teleportation and searching for a positive operator-valued measure~(POVM, i.e. the most general form of a quantum measurement) that Alice and Bob can implement to enhance performance. Our results show that the optimal POVM reduces to noiseless attenuation or amplification, depending on the operating regime of the teleportation protocol. For \ozlem{SR-DV} superdense coding, we likewise identify a POVM that enhance the mutual information between Alice and Bob beyond the classical channel capacity. We find that this advantage can also be achieved when Alice employs a simple noiseless attenuator or noiseless linear amplifier. In both teleportation and superdense coding, the optimisation problems are likely non-convex due to the wide parameter space, so the results reported here should be regarded as one possible solution. Nonetheless, it is encouraging that the NA and NLA circuits, being among the simplest POVMs that can be realised experimentally, already offer clear improvements over the baseline schemes.

Furthermore, our analysis reveals that not all maximally entangled Bell states are equally robust or useful for \ozlem{SR-DV} quantum communication. While some can be recovered up to small perturbations from the ideal case, other Bell states fail to reach the same fidelities. This distinction is known in the general two-qubit local-filtering or quasidistillation setting~\cite{horodecki1999general, verstraete2001local}. Here, we formally prove it for the specific lossy single-rail states considered in this work, showing that the choice of shared Bell state fundamentally affects teleportation performance and achievable fidelity.

This paper is organised as follows. In Sec.~\ref{sec:teleportation}, we present the details of the teleportation protocol, describing both the POVM and NLA/NA approaches together with their optimisation cost functions, and compare their performance in terms of the average fidelity achieved. In Sec.~\ref{sec:superdense_coding}, we outline the superdense coding protocol implemented with either POVMs or NLA/NA circuits, and assess their performance through the improvement obtained over the classical channel capacity. In Sec.~\ref{sec:conclusion}, we summarise our findings and present our conclusions.


\section{\label{sec:teleportation}SR-DV Teleportation}
In \ozlem{SR-DV} teleportation, a third party (commonly denoted as Charlie) prepares a maximally entangled Bell state of the form
\begin{equation}
    \label{eq:bell_state}
    \ket{\psi^+}=\frac{\ket{01}+\ket{10}}{\sqrt{2}}
\end{equation}
and transmits one qubit to Alice and the other to Bob through a pure-loss channel, in which photons can be lost to the environment. The shared state between Alice and Bob then becomes
\begin{equation}
\label{eq:lossy_bell_state}
\rho_{AB}\!=\!\tfrac{1}{2}\!
\begin{pmatrix}(1\!-\!T_A)+(1\!-\!T_B) & 0 & 0 & 0 \\
0 & T_B & \sqrt{T_AT_B} & 0 \\
0 & \sqrt{T_AT_B} & T_A & 0 \\
0 & 0 & 0 & 0
\end{pmatrix}\!,
\end{equation}
where $T_A$ and $T_B$ represent the transmission probabilities for Alice’s and Bob’s qubits through the pure-loss channel, respectively. 

Alice then interferes the unknown state $\ket{\phi}$ to be teleported with her share of the entangled pair and performs a Bell-state measurement. Depending on the measurement outcome, Bob applies the appropriate Pauli correction to his qubit: for outcome $00$, no operation is required; for $01$, he applies $X$ (bit-flip); for $10$, he applies $Z$ (phase-flip); and for $11$, he applies $ZX$ (bit-and-phase flip) gates, which are defined as
\begin{equation}
\label{eq:xgate}
    X=\begin{pmatrix}
        0 & 1 \\
        1 & 0
    \end{pmatrix},
\end{equation}
\begin{equation}
\label{eq:zgate}
    Z=\begin{pmatrix}
        1 & 0 \\
        0 & -1
    \end{pmatrix}.
\end{equation}

It is important to note that Charlie can also prepare other Bell states, such as $\ket{\phi^+} = (\ket{00} + \ket{11})/\sqrt{2}$. In the general two-qubit local-filtering/quasidistillation setting, different mixed-state families are not equally recoverable~\cite{horodecki1999general, verstraete2001local}. Here, for the specific lossy SR states considered in this work, when the distributed state is $\ket{\psi^+}$, the shared entanglement between Alice and Bob can be recovered through local operations and classical communication~(LOCC) up to small perturbations from the ideal case, achieving fidelities that approach but do not reach unity, albeit with a non-zero yet limited success probability in agreement with the general quasidistillation/local-filtering results of Refs.~\cite{horodecki1999general, verstraete2001local}. 
In contrast, when $\ket{\phi^+}$ is distributed, the entanglement cannot be perfectly recovered, even under arbitrarily small deviations from ideal the $\ket{\phi^+}$ state, as shown in Appendix~\ref{appendix:results_with_phi_plus} and proven in Appendix~\ref{appendix:proof_symmetric_loss}. Consequently, although the fidelities can still be improved through optimisation, they do not reach the same levels as in the $\ket{\psi^+}$ case. This behaviour highlights the direct connection between the strength of the shared entanglement and the fidelity of the teleported state.

\subsection{\label{sec:teleportation_POVM_optimisation}POVM Optimisation for SR-DV Teleportation}
\begin{figure}[t]
\includegraphics[scale=0.42]{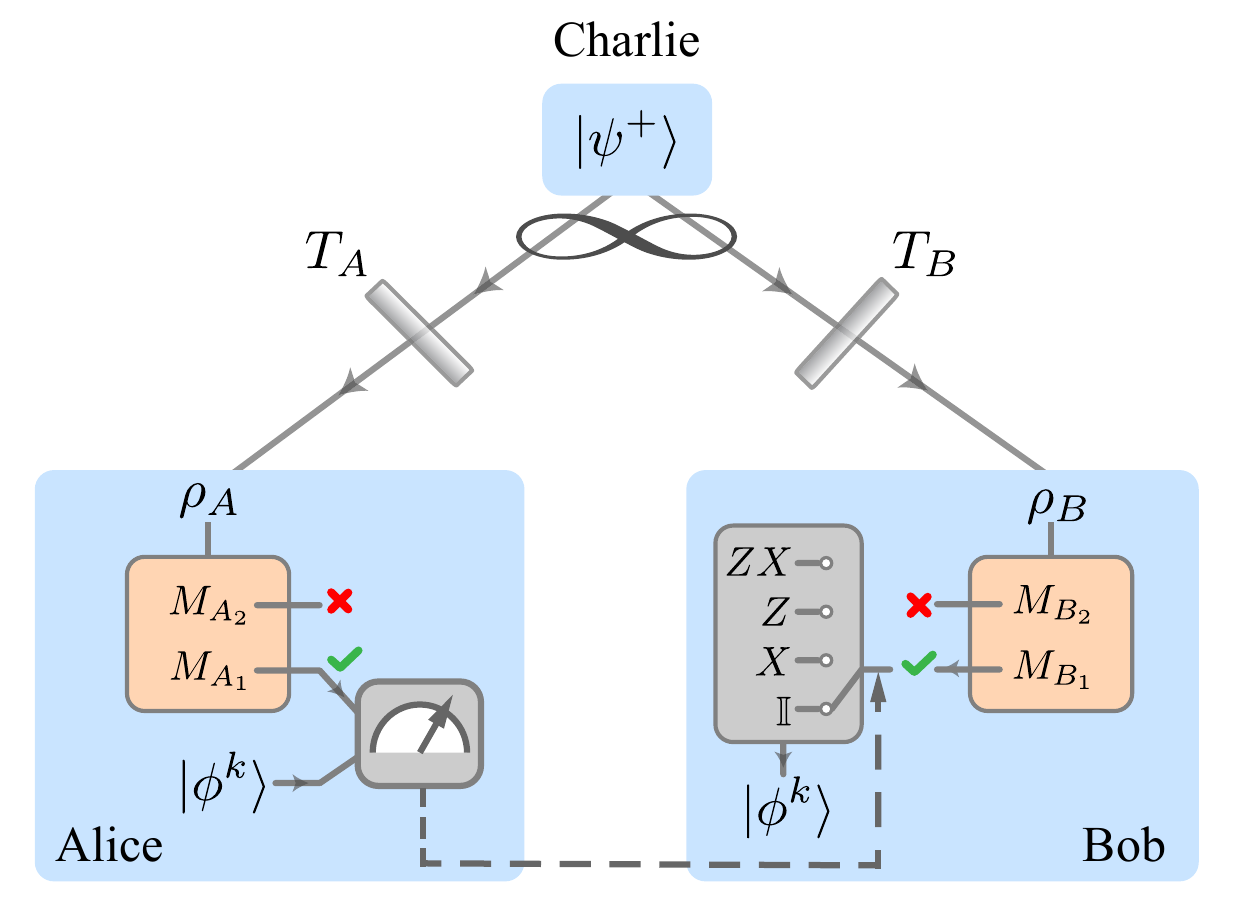}
\caption{\label{fig:figure1}Schematic of \ozlem{SR-DV} teleportation with the POVMs. Charlie prepares a maximally entangled Bell state and distributes one qubit each to Alice and Bob through pure-loss channels with transmittivities $T_A$ and $T_B$, respectively. Both parties apply a POVM to their received qubits. Upon a successful outcome, Alice combines the unknown input state with her qubit and performs a Bell-state measurement, sending the result to Bob, who applies the corresponding correction.
}
\end{figure}
Any transmission loss experienced by Alice’s and Bob’s qubits reduces the fidelity of the teleported state. To mitigate this effect, we propose that both parties apply a two-outcome POVM, as illustrated in Fig.~\ref{fig:figure1}, which effectively compensates for the channel losses. The corresponding measurement operators $M_{A_1}, M_{A_2}$ for Alice and $M_{B_1}, M_{B_2}$ for Bob are taken to be $2\times 2$ matrices with complex entries, and the associated POVM elements are defined as
\begin{equation}
    \label{eq:povm_elements}
    E_{X_m} = M_{X_m}^\dagger M_{X_m}, \quad X \in \{A,B\}, \; m \in \{1,2\}.
\end{equation}
For these to be valid POVMs, the elements need to satisfy the identity resolution 
\begin{equation}
    \label{eq:povm_identity_resolution}
    \sum_{m=1}^2 E_{X_m} = \mathbb{I}, \quad X \in \{A,B\}.
\end{equation}

Let the unknown input state to be teleported be denoted by $\rho_T=\ket{\phi}\!\bra{\phi}$, and Bob’s output state by $\rho_B$. Our goal is to maximise the fidelity $F(\rho_T,\rho_B)$ between the input state and Bob’s received state. However, optimising the POVMs with respect to a single input state would yield a state-dependent teleporter. To be universal, the teleporter needs to operate for any input state. For this reason, we randomly sample $N$ input states from the Bloch sphere and optimise the POVMs to maximise the average fidelity. These states take the form
\begin{equation}
\label{eq:random_states}
    \ket{\phi^k} = \cos\!\left(\tfrac{\theta_k}{2}\right)\ket{0} 
+ e^{i\phi_k}\sin\!\left(\tfrac{\theta_k}{2}\right)\ket{1},
\end{equation}
where $\theta_k=\arccos(2r_{1k}-1)$, $\phi_k=2\pi r_{2k}$, and 
$r_{1k},r_{2k}\sim\mathcal{U}[0,1]$.

The optimisation objective is then defined as the average fidelity over this ensemble of $N$ input states, maximised with respect to the POVM elements $M_{A_1}$ and $M_{B_1}$:
\begin{widetext}
\begin{equation}
    \label{eq:fidelity_maximisation}
    \Bar{F}=\max_{M_{A_1}, M_{B_2}}\;\frac{1}{N}\sum_{k=1}^{N}\!\bra{\phi^k}
    \frac{1}{P_{\psi^+}}(\Pi_{\psi^+}\otimes \mathbb{I}_2)
    \bigg(\rho_T^k \otimes 
    \frac{(M_{A_1}\otimes M_{B_1})\rho_{AB}(M_{A_1}\otimes M_{B_1})^\dagger}{P_{\text{succ}}}
    \bigg)(\Pi_{\psi^+}\otimes \mathbb{I}_2)^\dagger
    \ket{\phi^k},
\end{equation}
\end{widetext}
where $\ket{\phi^k}$ denotes the $k$-th input state sampled from the Bloch sphere as defined in Eq.~\eqref{eq:random_states}, 
$P_{\psi^+}$ is the success probability of the Bell-state projection with $\Pi_{\psi^+}=\ket{\psi^+}\!\bra{\psi^+}$ given in Eq.~\eqref{eq:bell_state}, and 
$P_{\text{succ}}$ is the success probability associated with the POVM elements applied by Alice and Bob. 
We emphasise that $P_{\mathrm{succ}}$ denotes the heralding probability of the local filtering step (POVM or NLA/NA), i.e., the fraction of trials kept after post-selection. In the low transmission regime where the fidelity gains are largest, the optimisation typically selects solutions at (or very close to) the imposed bound $P_{\mathrm{succ}} \approx 10^{-4}$, corresponding to a sacrifice ratio of order $1/P_{\mathrm{succ}} \sim 10^{4}$ trials per successful filtered event. The optimisation is carried out over the POVM elements $M_{A_1}$ and $M_{B_1}$ corresponding to the 
``success'' outcomes of Alice and Bob, subject to the following constraints:  

(i) the operators must define valid POVMs, that is
\begin{equation}
\label{eq:povm_constraints}
    E_X = M_X^\dagger M_X,\quad 0 \preceq E_X \preceq \mathbb{I}, \qquad X\in\{A,B\},
\end{equation}
with the complementary elements fixed by $E_{X_2}=\mathbb{I}-E_{X_1}$ and

(ii) the overall success probability satisfies
\begin{equation}
\label{eq:povm_prob_constraint}
    P_{\mathrm{succ}} \geq 10^{-4},
\end{equation}
to avoid unphysical solutions with vanishingly small probability of success.
\begin{figure}[t]
\includegraphics[scale=0.9]{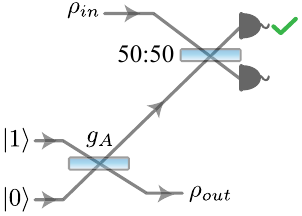}
\caption{\label{fig:figure2}Circuit used by Alice and Bob to implement NLA or NA. The beam splitter transmittivities $g_A$ and $g_B$ are tuned independently to maximise the mean fidelity of the teleported states. The input $\rho_{\text{in}}$ denotes the qubit (belonging to either Alice or Bob) entering the circuit for amplification or attenuation. For $g_X>0.5$ ($X\in{A,B}$), the corresponding qubit undergoes attenuation, while for $g_X<0.5$ it is amplified. A successful operation is heralded when exactly one detector registers a click, yielding the output state $\rho_{\text{out}}.$
}
\end{figure}
\begin{figure}[b]
\includegraphics[scale=0.42]{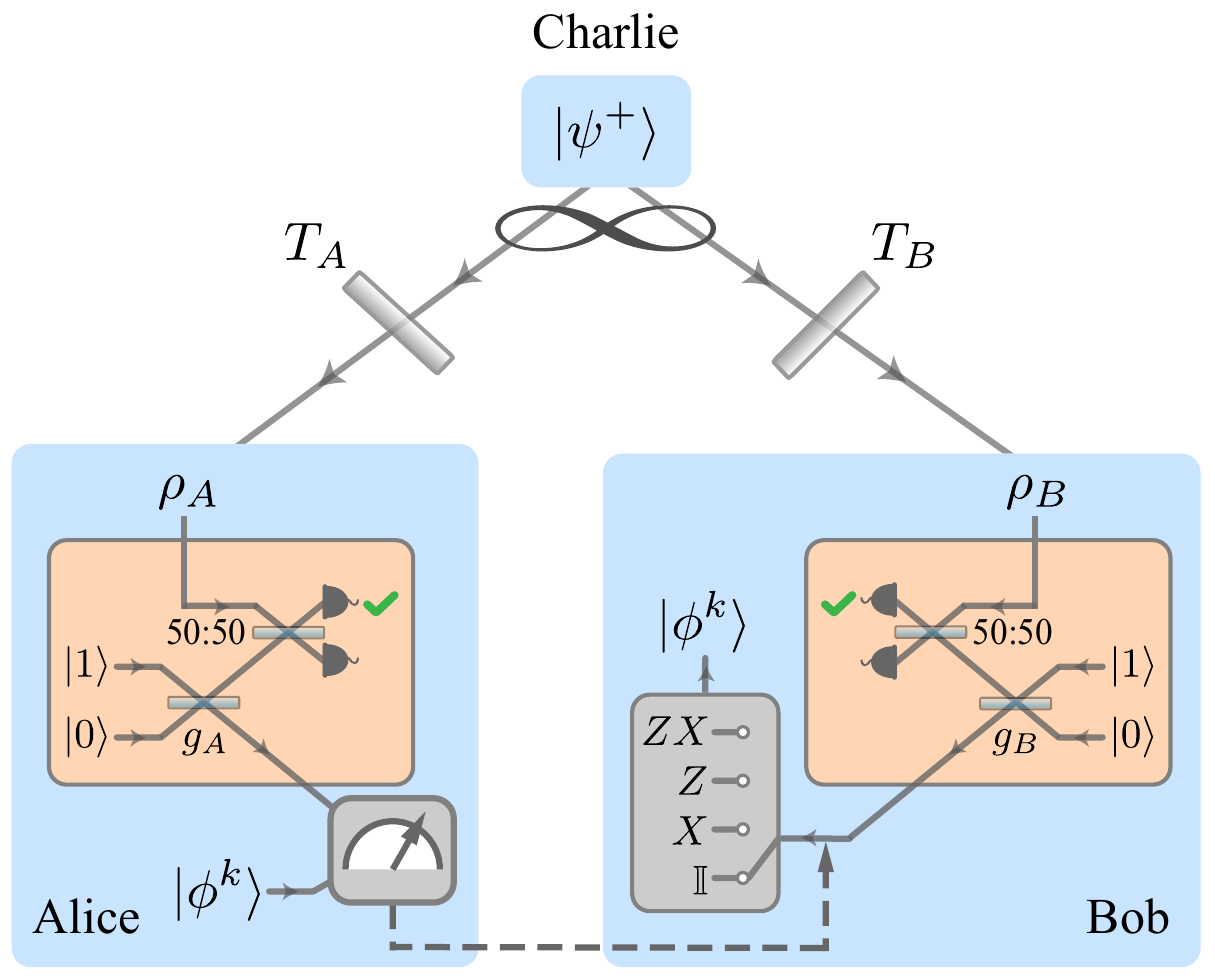}
\caption{\label{fig:figure3}Schematic of \ozlem{SR-DV} teleportation with the NLA/NA circuit. Charlie prepares a maximally entangled Bell state and distributes one qubit each to Alice and Bob through pure-loss channels with transmittivities $T_A$ and $T_B$, respectively. Each party applies an NLA/NA circuit to their received qubit. A successful operation is heralded by a single-photon detection. Alice then combines the unknown input state with her qubit and performs a Bell-state measurement, forwarding the result to Bob, who applies the appropriate correction.
}
\end{figure}

\subsection{\label{sec:teleportation_NLA_NA}Integration of Noiseless Attenuation and Noiseless Linear Amplification in SR-DV Teleportation}
In the previous section, we considered arbitrary two-outcome POVMs as a means to mitigate channel loss. We now turn to physically realisable operations with direct optical implementation such as noiseless linear amplification~(NLA) and noiseless attenuation~(NA). Instead of optimising over general measurement operators, we restrict the optimisation to the beamsplitter transmissivity of the NLA and NA operations to evaluate how these processes can enhance the fidelity of the \ozlem{SR-DV} teleportation. Both NLA and NA are implemented with the same circuit as shown in Fig.~\ref{fig:figure2}, where the input state is mixed with a superposition of vacuum and a single photon ancilla and heralded by a single photon detection. For the lossy state family derived from $\ket{\psi^+}$, this local NLA filtering is closely related to heralded qubit amplification which was demonstrated in Ref.~\cite{kocsis2013heralded}. This is because the relevant component of our SR entangled state is mathematically equivalent to a dual-rail single-photon (polarisation) qubit and is processed using mode-wise NLA stages. However, the role of the filter is different in our setting. Here, the NLA/NA is used as a local filter on a distributed entangled resource (rather than as a qubit-transmission amplifier), and we optimise and benchmark its effect on SR-DV teleportation fidelity and success probability.

A successful event occurs when exactly one detector clicks, in which case the input state is teleported to the output port and undergoes either amplification or attenuation, depending on the beamsplitter transmittivity, denoted $g_A$ and $g_B$ for Alice and Bob, respectively. When $g_{A}$ or $g_{B}$ exceeds 0.5 the corresponding qubit is attenuated, while values below 0.5 result in amplification. Figure~\ref{fig:figure3} illustrates how Alice and Bob apply the circuit. After Charlie distributes the maximally entangled state through pure-loss channels with transmittances $T_A$ and $T_B$, Alice and Bob each apply this circuit to their respective qubits. At this stage, the unnormalised shared state between them can be written as
\begin{equation}
    \label{eq:general_state_AB}
    \Tilde{\rho}_{AB}=\begin{pmatrix}
        a & 0 & 0 & 0 \\
        0 & b & c & 0 \\
        0 & c & d & 0 \\
       0 & 0 & 0 & 0   
    \end{pmatrix},
\end{equation}
with the matrix entries
\begin{equation}
\label{eq:general_state_entry_1}
a=0.125g_{B}(1-g_{A})\big((1-T_A)+(1-T_B)\big),  
\end{equation}
\begin{equation}
    \label{eq:general_state_entry_2}
    b=0.125(1-g_A)(1-g_B)T_B,
\end{equation}
\begin{equation}
    \label{eq:general_state_entry_3}
    c = 0.125\sqrt{g_Ag_B(1-g_A)(1-g_B)T_AT_B},
\end{equation}
\begin{equation}
    \label{eq:general_state_entry_4}
    d = 0.125g_Ag_BT_A,
\end{equation}
Here $g_A$ and $g_B$ denote the beamsplitter transmittivity of Alice’s and Bob’s circuits, respectively. The success probability of the circuit is given as $P_{\text{succ}}=2\text{Tr}[\Tilde{\rho}_{AB}]$ and the normalised state becomes $\Tilde{\rho}_{AB}^n=\Tilde{\rho}_{AB}/\text{Tr}[\Tilde{\rho}_{AB}]$.
\begin{figure*}[t]
\includegraphics[scale=0.45]{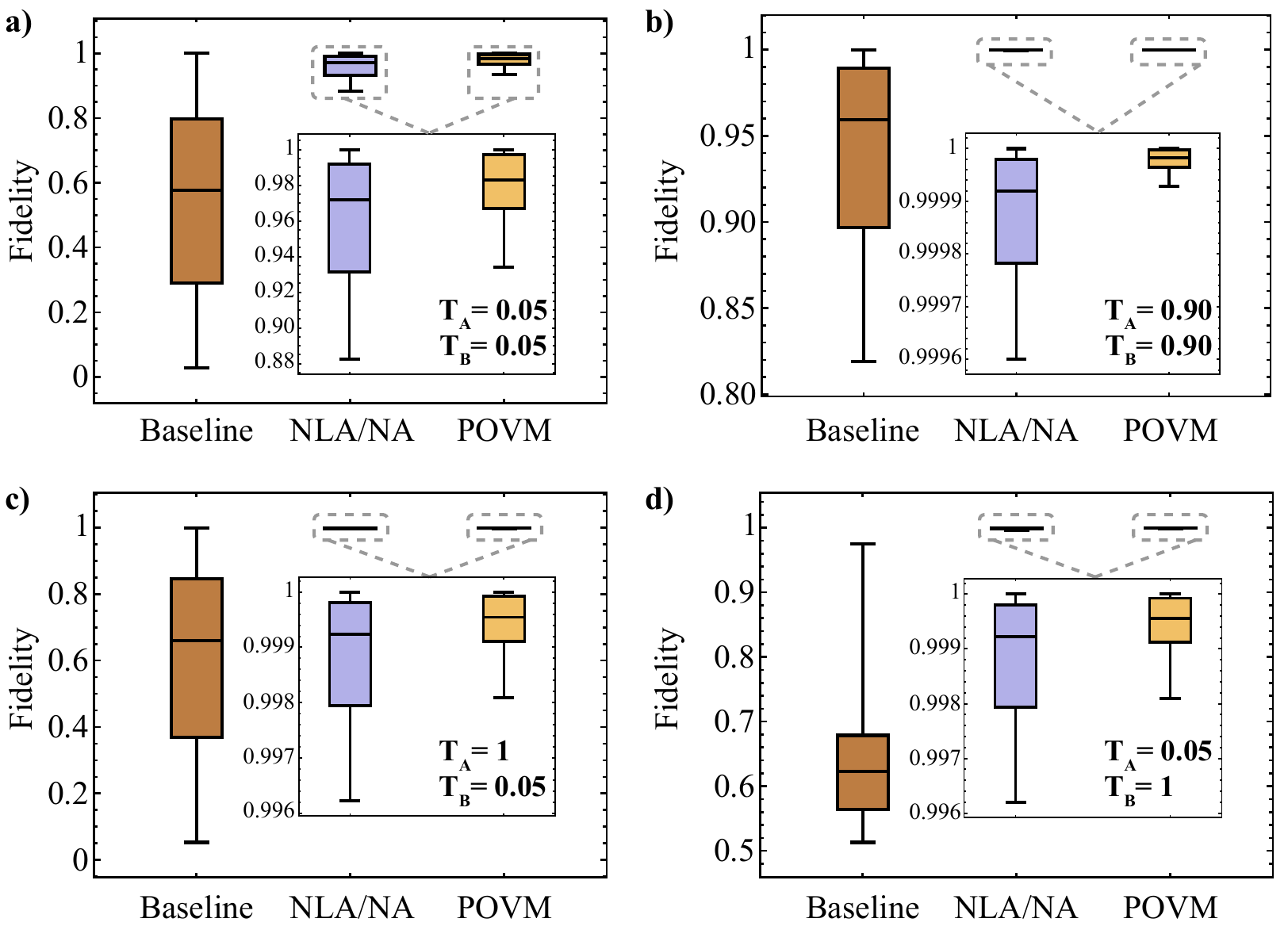}
\caption{\label{fig:figure4}Fidelities of \ozlem{SR-DV} teleportation for $N=300$ input states. Brown boxes show the baseline performance without any additional operation. Blue boxes correspond to teleportation where Alice and Bob apply NLA or NA, with the beamsplitter transmittivities optimised to maximise the average fidelity. Orange boxes correspond to teleportation where Alice and Bob implement a POVM optimised for the best average fidelity. \textbf{(a)} Charlie distributes qubits to Alice and Bob with $T_A = T_B = 0.05$.  
\textbf{(b)} Charlie distributes qubits to Alice and Bob with $T_A = T_B = 0.9$. \textbf{(c)} Alice receives her qubit without loss ($T_A = 1$), while Bob’s qubit experiences a transmission of $T_B = 0.05$. \textbf{(d)} Bob receives his qubit without loss ($T_B = 1$), while Alice’s qubit experiences a transmission of $T_A = 0.05$.}
\end{figure*}
\begin{figure*}[t]
\includegraphics[scale=0.46]{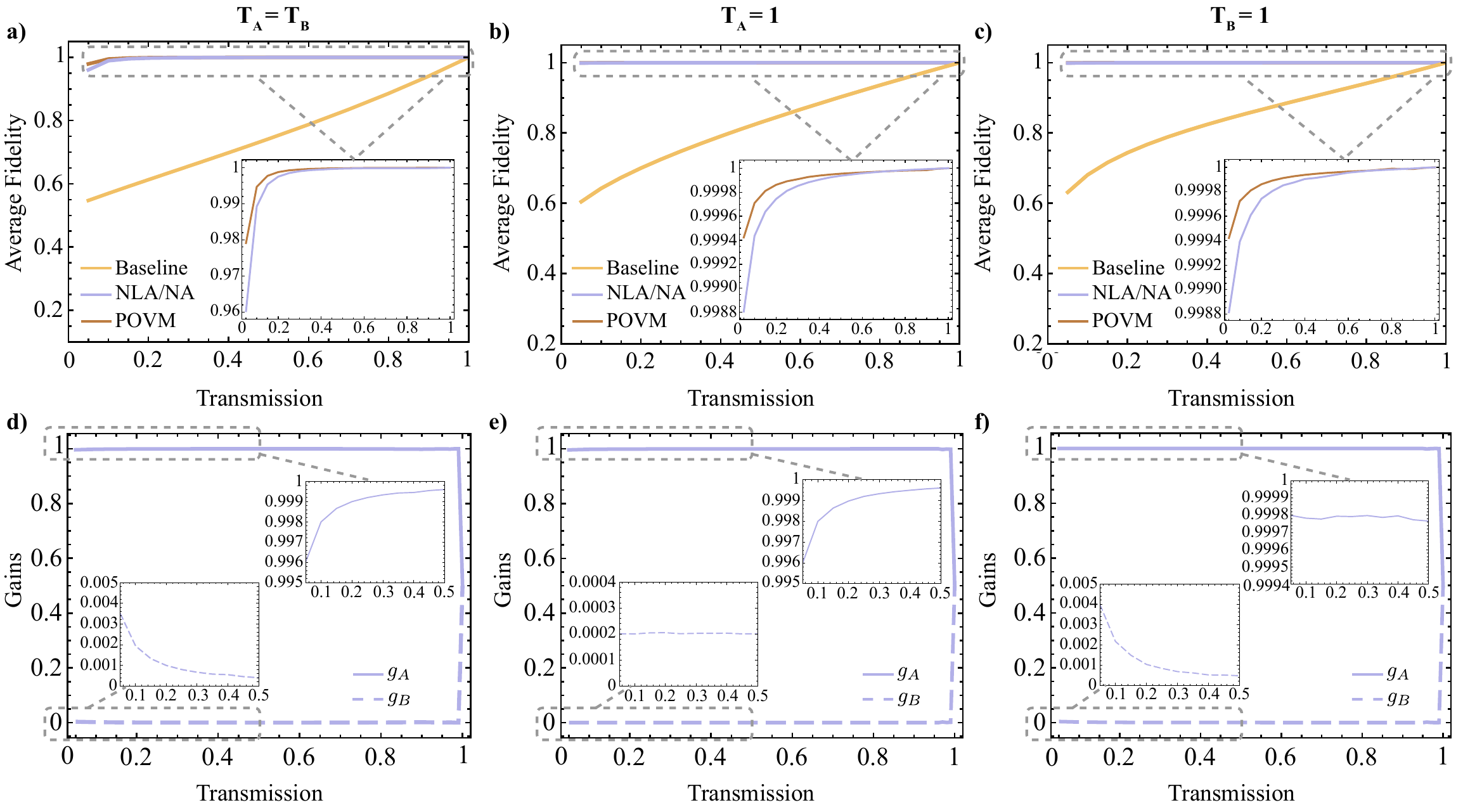}
\caption{\label{fig:figure5}Comparison of baseline, NLA/NA, and POVM teleportation across different transmission regimes, shown in terms of average fidelities and optimised transmittivities. \textbf{(a)} Average fidelities for the three cases: baseline teleportation (orange), teleportation with NLA/NA using optimised transmittivities (blue), and teleportation with optimised POVMs (brown). In this panel, Charlie distributes the qubits to Alice and Bob through pure-loss channels with equal transmissions $T_A = T_B$. Panels \textbf{(b)} and \textbf{(c)} present the fidelities under asymmetric transmissions. In panel \textbf{(b)}, Alice’s qubit is transmitted without loss ($T_A = 1$), while Bob’s qubit has transmission $T_B$ varying from $0$ to $1$. Conversely, in panel \textbf{(c)}, Bob’s qubit is transmitted without loss ($T_B = 1$), while Alice’s qubit has transmission $T_A$. Panels \textbf{(d)}, \textbf{(e)}, and \textbf{(f)} show the optimised transmittivities of NLA/NA teleportation corresponding to the regimes $T_A = T_B$, $T_A = 1$, and $T_B = 1$, respectively.
}
\end{figure*}

Similar to the POVM optimisation, we optimise the circuit transmittivities to maximise the average fidelity over an ensemble of $N$ input states. The objective function is given by
\begin{widetext}
\begin{equation}
    \label{eq:fidelity_maximisation_nla_na}
    \Bar{F}=\max_{g_A,g_B}\;\frac{1}{N}\sum_{k=1}^{N}\!\bra{\phi^k}
    \frac{1}{P_{\psi^+}}(\Pi_{\psi^+}\otimes \mathbb{I}_2)
    \big(\rho_T^k \otimes 
    \Tilde{\rho}_{AB}^n
    \big)(\Pi_{\psi^+}\otimes \mathbb{I}_2)^\dagger
    \ket{\phi^k},
\end{equation}
\end{widetext}
subject to constraints $0\leq\!g_A\!\leq1$, $0\leq\!g_B\!\leq1$ and $P_{\mathrm{succ}} \geq 10^{-4}$. Here, $P_{\mathrm{succ}}$ is the heralding probability of the local filter. In the low transmission regime, the optimal solutions typically occur at (or very close to) $P_{\mathrm{succ}}\approx 10^{-4}$, giving a sacrifice ratio of $1/P_{\mathrm{succ}}\sim 10^{4}$.
In this optimisation, we assume that Alice obtains the Bell-state measurement outcome $00$, in which case Bob does not need to apply any correction to his qubit. Bob's state after Alice's measurement and feedforward is expressed as
\begin{equation}
\label{eq:bobs_state}
\Tilde{\rho}_B=\text{Tr}_{12}\bigg[\frac{1}{P_{\psi^+}}(\Pi_{\psi^+}\otimes \mathbb{I}_2)\big(\rho_T^k \otimes 
    \Tilde{\rho}_{AB}^n
    \big)(\Pi_{\psi^+}\otimes \mathbb{I}_2)^\dagger\bigg],
\end{equation}
where $\text{Tr}_{12}$ denotes the partial trace over the input state to be teleported and Alice’s mode.

\subsection{\label{eq:comparison_of_results_teleporter}Comparison of Teleportation Schemes}
Figure~\ref{fig:figure4} illustrates the improvement in average teleportation fidelity achieved by optimised NLA/NA and POVM operations over $N=300$ input states, relative to the baseline. In Fig.~\ref{fig:figure4}(a), where both qubits undergo a transmission of $T_A = T_B = 0.05$, the baseline teleportation fidelities range from 0.03 to 0.9998, with a median of $0.58$. When Alice and Bob apply either NLA or NA, determined by the beamsplitter transmissivity in Fig.~\ref{fig:figure2}, the teleportation fidelities with respect to the input state improve, yielding a median of 0.97. 
The minimum fidelities increase substantially, and the overall distribution becomes narrower, with values ranging from 0.88 to 1. In this case, the optimised transmittivities are $g_A = 0.9996$ and $g_B = 0.0035$, corresponding to Alice applying noiseless attenuation and Bob applying noiseless linear amplification which together lead to an overall success probability of $P_{\text{succ}} \approx 10^{-4}$. When Alice and Bob apply an optimised POVM instead of the NLA/NA circuit, the teleportation fidelities lie within a narrower range of $0.93–1$ with a median of 0.98, achieving improvements comparable to the NLA/NA case, with an overall success probability of $P_{\text{succ}} \approx 10^{-4}$.

In Fig.~\ref{fig:figure4}(b), where both qubits are transmitted with $T_A = T_B = 0.9$, the application of NLA/NA or a POVM again enhances the teleportation fidelities. 
The improvements are even more pronounced in the case of $T_A = T_B = 0.9$, as the lower losses in this regime lead to higher fidelities, reflected by the tighter 25\% and 75\% quartiles of the baseline teleportation. Both NLA/NA and POVM operations narrow the distribution, obtaining a more consistent performance with a median fidelity of $\approx 0.9999$, and success probabilities of $P_{\text{succ}} \approx 8\times10^{-4}$ for NLA/NA and $P_{\text{succ}} \approx 2\times10^{-4}$ for the POVM. In this regime, the optimised transmittivities are $g_A = 0.9982$ and $g_B = 0.0018$, corresponding to Alice applying an NA and Bob applying an NLA. When the losses are asymmetric between Alice and Bob, the optimisation results follow the same pattern as the symmetric case as shown in Fig.~\ref{fig:figure4} in panels (c) and (d). In panel (c), where $T_A = 1$ and $T_B = 0.05$, the baseline fidelities span from $0.05$ to $0.99$ with a median fidelity of $F=0.66$. In contrast, panel (d), corresponding to $T_A = 0.05$ and $T_B = 1$, exhibits baseline fidelities in the range of $0.51$ to $0.97$ with median fidelity of $F=0.62$. Both NLA/NA and POVM operations enhance the fidelities, narrowing their spread to approximately $0.996$--$1$, with comparable success probabilities. In panels (c) and (d), Alice and Bob employ the same configuration as in the symmetric case, Alice applies an NA and Bob an NLA, with transmittivity values similar to those used in panels (a) and (b).

Figure~\ref{fig:figure5} further illustrates how both optimised POVM and NLA/NA operations enhance the average teleportation fidelities. In Fig.~\ref{fig:figure5}(a), Alice’s and Bob’s qubits undergo the same transmission before teleportation. Initially, the average fidelity achieved by the optimised POVMs is slightly higher than that of the teleporter employing the NLA/NA operation. The difference, however, is minor, with $\bar{F}=0.98$ for the optimised POVMs and $\bar{F}=0.96$ for the NLA/NA scheme, compared to the baseline teleportation fidelity of $\bar{F}=0.55$. This corresponds to an improvement of approximately $75\%$ and $78\%$, respectively, for the case of $T_A = T_B = 0.05$. As the channel transmittivities increase, the performances of the optimised POVM and NLA/NA schemes converge, coinciding around $T_A = T_B = 0.5$. In this regime, both achieve an average fidelity of $\bar{F} \approx 0.9999$, while the baseline fidelity remains at $\bar{F} \approx 0.74$. Similarly, in Figs.~\ref{fig:figure5}(b) and (c), where only Alice’s or Bob’s qubit is subject to transmission loss, both operations give fidelities approaching unity across all transmission values. The optimised POVMs provide a slight advantage over NLA/NA with optimised transmittivities, however, the difference is marginal ($0.06\%$). For $T_A=1, T_B = 0.05$ or $T_A = 0.05, T_B=1$, the average fidelity increases by approximately $66\%$ and $59\%$, respectively, compared to the baseline. In general, a slight advantage of the optimised POVMs over NLA/NA is expected because the POVM optimisation ranges over the full set of single-qubit non-unitary Kraus operators $M$, whereas the NLA/NA scheme considered here is restricted to an experimentally motivated one-parameter family that is diagonal in the single-rail basis. \ozlemREV{In Appendix~\ref{appendix:POVM_NLA_Matrices}, we provide explicit examples of the optimised POVMs and find that their dominant action is already largely diagonal in the single-rail basis. Motivated by this, we also repeated the optimisation over general local diagonal filters and found that these closely reproduce the full POVM results, while the constrained NLA/NA family falls slightly below both in the low-transmission regime when all optimisations are performed under the same success probability threshold, $P_{\mathrm{succ}} \ge 10^{-4}$. This indicates that the small gap is not primarily due to off-diagonal or unitary freedom, but rather reflects the more restrictive structure of the constrained NLA/NA family within the broader class of diagonal filters. A related benchmark based on directly optimising the conditional Bell-state fidelity $F_{\Psi^+}$ is presented in Appendix~\ref{appendix:optimisation_merit_psip} and shows the same qualitative behaviour. We also find that, when the NLA/NA threshold is relaxed to $P_{\mathrm{succ}} \ge 5 \times 10^{-5}$, the NLA/NA results become comparable to, and closely follow, the POVM and diagonal-filter curves in all regimes. In practice, the gap remains small, and} we therefore focus on NLA/NA as the more practical option, since it has established physical implementations, whereas a general optimal POVM would require a dedicated realisation of the corresponding non-unitary filter.

Figures~\ref{fig:figure5}(d)--(f) show the optimised NLA/NA transmittivity values that maximise the average fidelity across different transmission regimes. The optimal strategy remains consistent across all cases: Alice applies a noiseless attenuation operation while Bob applies a noiseless linear amplification operation, each with a success probability of approximately $P_\text{succ} \approx 10^{-4}$. In all cases, Alice’s gains are generally above $g_A > 0.5$ and Bob’s below $g_B < 0.5$, reflecting their respective attenuation and amplification roles. As the transmission increases and approaches $T_A = T_B = 1$, both transmittivities converge toward $g_A = g_B = 0.5$, corresponding to the lossless limit where no operation is required.


\section{\label{sec:superdense_coding}SR-DV Superdense Coding}
In superdense coding, as in \ozlem{SR-DV} teleportation, a third party (Charlie) distributes an entangled state. In the absence of loss, this is the maximally entangled Bell state defined in Eq.~\eqref{eq:bell_state}. When transmission losses are present, however, the shared resource degrades, and Charlie instead distributes the following state, sending one qubit to Alice and the other to Bob.
\begin{equation}
    \label{eq:bell_state_non_maximal}
    \ket{\psi_C}\!\!(p)=\sqrt{p}\ket{01}+\sqrt{1-p}\ket{10},
\end{equation}
where $p \in [0,1]$ parametrises the degree of entanglement of the shared state: $p=0.5$ corresponds to the maximally entangled Bell state $\ket{\psi^+}$, while $p \to 0$ or $p \to 1$ gives a separable state. Charlie then distributes one qubit to Alice and the other to Bob through pure-loss channels with transmissions $T_A$ and $T_B$, after which the shared state between Alice and Bob becomes
\begin{widetext}
\begin{equation}
\label{eq:lossy_entangled_state_superdense}
\rho_{AB}(p)=
\begin{pmatrix}(1-p)(1-T_A)+p\;(1-T_B) & 0 & 0 & 0 \\
0 & p\;T_b & \sqrt{p\;(1-p)\;T_A\;T_B} & 0 \\
0 & \sqrt{p\;(1-p)\;T_A\;T_B} & p\;T_A & 0 \\
0 & 0 & 0 & 0
\end{pmatrix}\!.
\end{equation}    
\end{widetext}

Depending on the classical message Alice wishes to transmit, she applies a corresponding quantum operation to her qubit. Specifically, to send $00$ she applies no operation, for $01$ she applies the Pauli-$X$ gate, for $10$ the Pauli-$Z$ gate, and for $11$ the combined operation $ZX$, the shared state between Alice and Bob after Alice's encoding can be expressed as
\begin{equation}
    \label{eq:Alices_encoded_state_superdense}
    \rho_{AB}(p,\sigma_k)=(\sigma_k\otimes\mathbb{I}_2)\rho_{AB}(p)(\sigma_k\otimes\mathbb{I}_2)^\dagger,
\end{equation}
where $\sigma_k \in \{\mathbb{I}, X, Z, ZX\}$ denotes the Pauli operation applied by Alice to encode her classical information on the qubit. She then transmits the encoded qubit through another pure-loss channel to Bob, who decodes the message by performing a Bell-state measurement on the two qubits. The resulting two-qubit state, after Alice’s encoded qubit is transmitted to Bob through a pure-loss channel with transmittivity $T_f$, can be written as
\begin{align}
    \rho_{AB}(p,\sigma_k,T_f)=&\text{Tr}_1\big[\big(\text{BS}(T_f)\otimes\mathbb{I}_2\big)\big(\ketbra{0}{0}\otimes\rho_{AB}(p,\sigma_k)\big)\nonumber \\
    &\big(\text{BS}(T_f)\otimes\mathbb{I}_2\big)^\dagger\big].
\end{align}
Here, the pure-loss channel is modelled with a beamsplitter denoted as $\text{BS}$ with a transmissivity \text{$T_f$}, where the beamsplitter mixes the input mode with the vacuum. The beamsplitter transformation can be defined as
\begin{equation}
\text{BS}(T_f)=\mathrm{exp}[\mathrm{cos}^{-1}({\sqrt{T_f}})(\hat{a}^\dagger\hat{b}-\hat{a}\hat{b}^\dagger)],
\label{eq:beamsplitter}
\end{equation}
\text{$\hat{a}$} and \text{$\hat{b}$} are the annihilation operators, while \text{$\hat{a}^\dagger$} and \text{$\hat{b}^\dagger$} are the creation operators of the two modes, respectively. In the absence of loss between Alice and Bob, and assuming Charlie has distributed the maximally entangled state, the mutual information shared between Alice and Bob is given by
\begin{equation}
    \label{eq:mutual_info_maximal}
    I_{AB}=\log_2(\text{dim})+S(\rho_B)-S(\rho_{AB}),
\end{equation}
where dim denotes the Hilbert-space dimension, which is $2$ for a qubit system, while $S(\rho_B)$ and $S(\rho_{AB})$ are the von Neumann entropies of Bob’s reduced state and the joint state of Alice and Bob, respectively. When both Alice’s and Bob’s qubits are subject to transmission losses, and Alice’s encoded qubit is further sent through a pure-loss channel, the mutual information between Alice and Bob is quantified using the Holevo bound~\cite{holevo1998capacity}, which represents the maximum classical information extractable from a quantum channel. In this case, the mutual information is expressed as
\begin{align}
    \label{eq:mutual_information_quantum}
    I_{AB}=&\max_{p,p_k}\bigg[S\bigg(\sum_{k=1}^4p_k\rho_{AB}(p,\sigma_k,T_f)\bigg)\nonumber \\
    &-\sum_{k=1}^4p_kS\big(\rho_{AB}(p,\sigma_k,T_f)\big)\bigg],
\end{align}
where $p_k$ is the probability that Alice prepares and transmits the state encoded with $\sigma_k$. $\rho_{AB_k}$ is the corresponding shared state after this encoding and transmission, and $S(\cdot)$ denotes the von Neumann entropy.
\begin{figure}[b]
\includegraphics[scale=0.42]{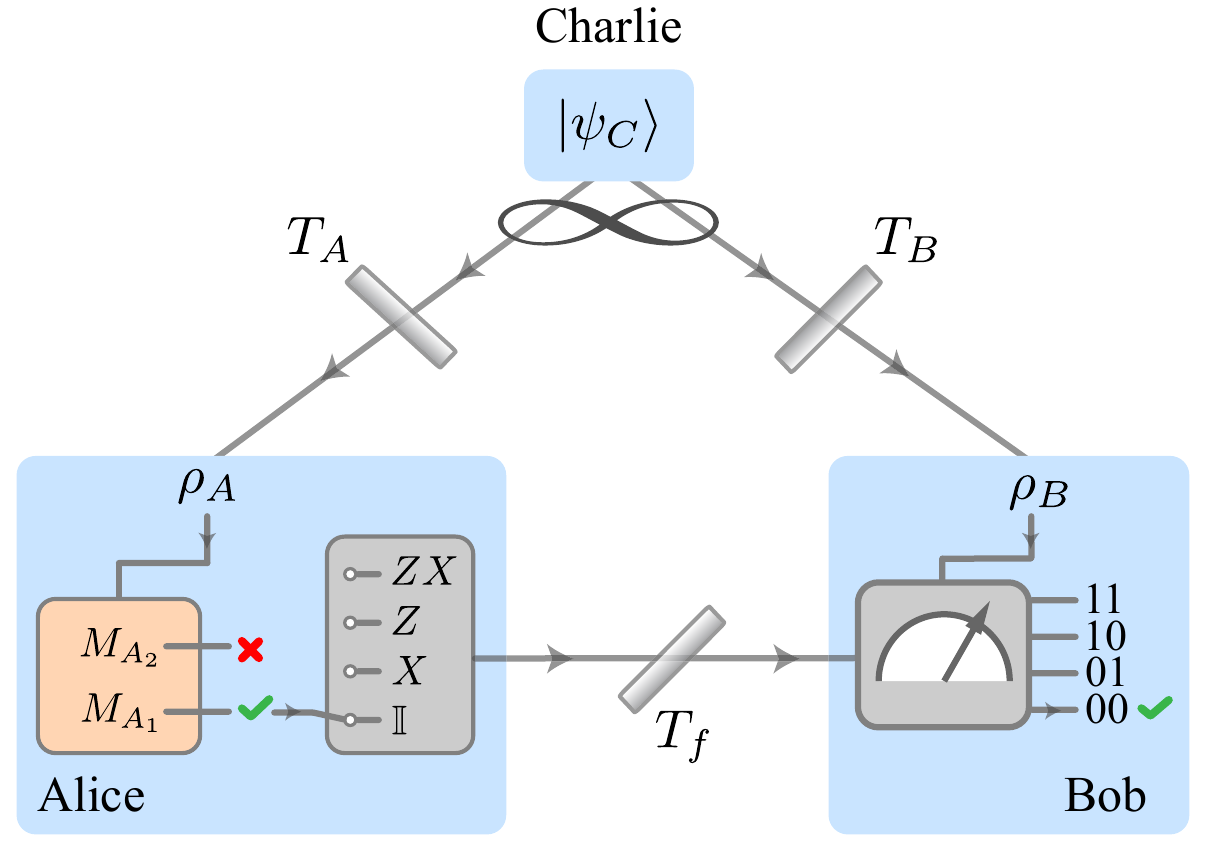}
\caption{\label{fig:figure6}Schematic of the superdense coding with the POVM. Charlie prepares an entangled state and distributes one qubit each to Alice and Bob through pure-loss channels with transmittivities $T_A$ and $T_B$, respectively. Alice first applies a POVM to her received qubit. Upon a successful outcome, she encodes her message using one of the Pauli gates and transmits the qubit to Bob through a pure-loss channel with transmittivity $T_f$. Bob then performs a Bell-state measurement on his qubit and the one received from Alice, thereby decoding the message. 
}
\end{figure}

If Charlie prepares a maximally entangled state and distributes it over lossless channels, the maximum tolerable loss in superdense coding, when Alice sends her encoded qubit to Bob, is $T_f = 0.5$. This result is derived under the assumption that in the absence of entanglement between Alice and Bob, i.e., in the classical case where Alice simply prepares and sends a single qubit to Bob, the transmitted information is always 1 bit. However, this comparison assumes that the single qubit in the classical case is transmitted without any loss, which is not a fair basis for evaluating performance.

In the classical case, let us assume that Alice prepares a qubit in the following form
\begin{equation}
    \label{eq:classical_case}
    \ket{\chi_\pm}\!(\eta)=\sqrt{\eta}\ket{0}\pm\sqrt{1-\eta}\ket{1},
\end{equation}
where $\eta \in [0,1]$ parametrises the relative weighting of $\ket{0}$ and $\ket{1}$ in the superposition. Alice sends $\ket{\chi_+}$ and $\ket{\chi_-}$ with probabilities $p_c$ and $1-p_c$, respectively, and both states are transmitted through a pure-loss channel with transmittivity $T_f$. The state that Bob receives can be expressed as
\begin{equation}
    \label{eq:rho_capacity_state}
    \rho_{B_\pm}\!(\eta)\!=\!\text{Tr}_1[(\text{BS}(T_f)\otimes\mathbb{I}_2)(\ketbra{0}{0}\otimes \ketbra{\chi_\pm}{\chi_\pm})(\text{BS}(T_f)\otimes\mathbb{I}_2)^\dagger].
\end{equation}
Then the classical capacity of the channel as a function of the pure-loss channel with transmittivity of $T_f$ when a single qubit is used expressed as
\begin{align}
    \label{eq:capacity_single_qubit}
    C(T_f)=&\max_{\eta,p_c}\bigg[S\big(p_c\rho_{B_+}(\eta)+(1-p_c)\rho_{B_-}(\eta)\big)\nonumber\\
    &-\big[p_cS\big(\rho_{B_+}(\eta)\big)+(1-p_c)S\big(\rho_{B_-}(\eta)\big)\big]\bigg],
\end{align}
subject to the constraints $0 \leq \eta \leq 1$ and $0 \leq p_c \leq 1$.

\subsection{\label{sec:superdense_POVM_optimisation}POVM Optimisation for Superdense Coding}
After both Alice and Bob receive their qubits, Alice applies a POVM as shown in Fig.~\ref{fig:figure6}, after a successful outcome, she then encodes the classical bits using one of the Pauli gates and sends the encoded qubit through a pure-loss channel. We apply the POVM only on Alice’s side and condition the analysis on successful outcomes. In this way, failed events correspond to Alice not sending a state and are excluded from the statistics. If Bob were to apply a POVM, however, both parties would need to succeed simultaneously, and the overall success probability would explicitly reduce the achievable information rate. To this end, the state between Alice and Bob after Alice's POVM can be expressed as
\begin{equation}
    \label{eq:rhoab_superdense}
    \Tilde{\rho}_{AB}(p,M_{A_1})=\frac{(M_{A_1}\otimes\mathbb{I}_2)\rho_{AB}(p)(M_{A_1}\otimes\mathbb{I}_2)^\dagger}{P_\text{succ}},
\end{equation}
where $P_{\text{succ}}$ is the success probability associated with the POVM elements applied by Alice and the optimisation is carried out over the POVM element $M_{A_1}$ corresponding to the 
``success'' outcome of Alice subject to the same constraints given in Eqs.~\eqref{eq:povm_constraints} and~\eqref{eq:povm_prob_constraint}. After a successful POVM outcome, Alice encodes her classical information by applying one of the Pauli gates to her qubit and then transmits it through the channel where the shared state between Alice and Bob can be expressed as
\begin{align}
    \label{eq:rhoab_before_measurement_superdense}
    \Tilde{\rho}_{AB}(p,M_{A_1},\sigma_k,T_f)\!=&(\text{BS}(T_f)\otimes\mathbb{I}_2)(\sigma_k\otimes\mathbb{I}_2)\Tilde{\rho}_{AB}(p,M_{A_1})\nonumber \\
    &(\sigma_k\otimes\mathbb{I}_2)^\dagger(\text{BS}(T_f)\otimes\mathbb{I}_2)^\dagger,
\end{align}
where $\sigma_k \in \{\mathbb{I}, X, Z, ZX\}$ denotes the Pauli operation applied by Alice to encode her classical information on the qubit, and $T_f$ represents the transmission channel through which the qubit is sent to Bob, modelled using Eq.~\eqref{eq:beamsplitter}. As noted earlier, Bob decodes the message by performing a Bell-state measurement on the qubit received from Alice together with his own. The mutual information between Alice and Bob is then optimised over both the probabilities of preparing each encoded state and the parameters of Alice’s POVM as
\begin{align}
    \label{eq:mutual_info_maximum_superdense}
    I_{AB}\!=&\!\max_{p,p_k,M_{A_1}}\bigg[S\bigg(\sum_{k=1}^4p_k\Tilde{\rho}_{AB}(p,M_{A_1},\sigma_k,T_f)\bigg)\nonumber \\
    &-\sum_{k=1}^4p_kS\big(\Tilde{\rho}_{AB}(p,M_{A_1},\sigma_k,T_f)\big)\bigg].
\end{align}

Superdense coding provides an advantage only if the number of classical bits transmitted to Bob exceeds what could be achieved by sending a single qubit through the same lossy channel. To quantify this, we adopt the notion of \textit{quantum advantage}, proposed by Samanta~\cite{samanta2024continuous}~\textit{et al.}, defined as the difference between the maximum mutual information achievable between Alice and Bob using entanglement and the classical channel capacity when only a single qubit is transmitted. Based on this definition, our cost function for the optimisation is given by
\begin{figure}[t]
\includegraphics[scale=0.42]{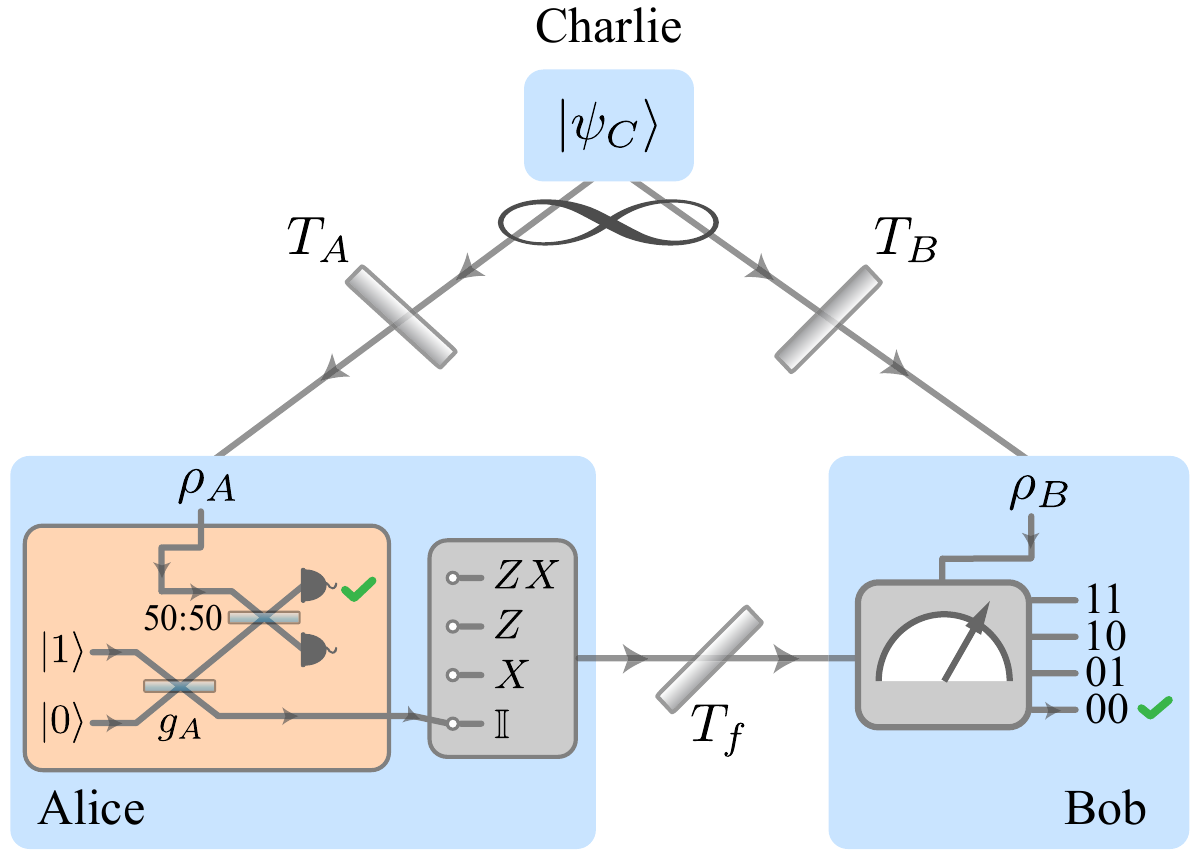}
\caption{\label{fig:figure7}Schematic of the superdense coding protocol with the NLA/NA circuit. Charlie generates an entangled state and distributes one qubit each to Alice and Bob through pure-loss channels with transmittivities $T_A$ and $T_B$. Alice applies the NLA/NA circuit to her qubit; upon a detector click, the state is either amplified or attenuated depending on the transmittivity parameter $g_A$. Following a successful click, she encodes her message using a Pauli gate and sends the qubit to Bob through another pure-loss channel with transmittivity $T_f$. Bob then performs a Bell-state measurement on his qubit and the one received from Alice to decode the message.
}
\end{figure}
\begin{align}
    \label{eq:Qad_POVM_superdense}
    Q_{\text{POVM}}(T_f)=&\!\max_{p,p_k,M_{A_1}}\bigg[S\bigg(\sum_{k=1}^4p_k\Tilde{\rho}_{AB}(p,M_{A_1},\sigma_k,T_f)\bigg)\nonumber \\
    &-\sum_{k=1}^4p_kS\big(\Tilde{\rho}_{AB}(p,M_{A_1},\sigma_k,T_f)\big)\bigg]-C(T_f),
\end{align}
where $C(T_f)$ denotes the classical channel capacity, given in Eq.~\eqref{eq:capacity_single_qubit}, which must be optimised beforehand as a prerequisite to the quantum advantage optimisation. The optimisation in Eq.~\eqref{eq:Qad_POVM_superdense} is subject to the same constraints as in Eqs.~\eqref{eq:povm_constraints} and~\eqref{eq:povm_prob_constraint}, with the additional conditions
\begin{equation}
\label{eq:superdense_povm_constraint}
    0 \leq p \leq 1, 
    \qquad \sum_{k=1}^{4} p_k = 1, 
    \qquad 0 \leq p_k \leq 1 \ \ \forall k.
\end{equation}

\subsection{\label{sec:superdense_NLA_optimisation}Integration of Noiseless Attenuation and Noiseless Linear Amplification in Superdense Coding}
Similar to the teleportation scheme, we now assume that Alice employs a physical NA or NLA using the circuit shown in Fig.~\ref{fig:figure7}, rather than a two-outcome POVM, to mitigate channel losses. After Charlie distributes the qubits through lossy channels, the shared state between Alice and Bob is given in Eq.~\eqref{eq:lossy_entangled_state_superdense}. Alice then applies either the NA or NLA, leading to the following unnormalised shared state
\begin{figure*}[t]
\includegraphics[scale=0.46]{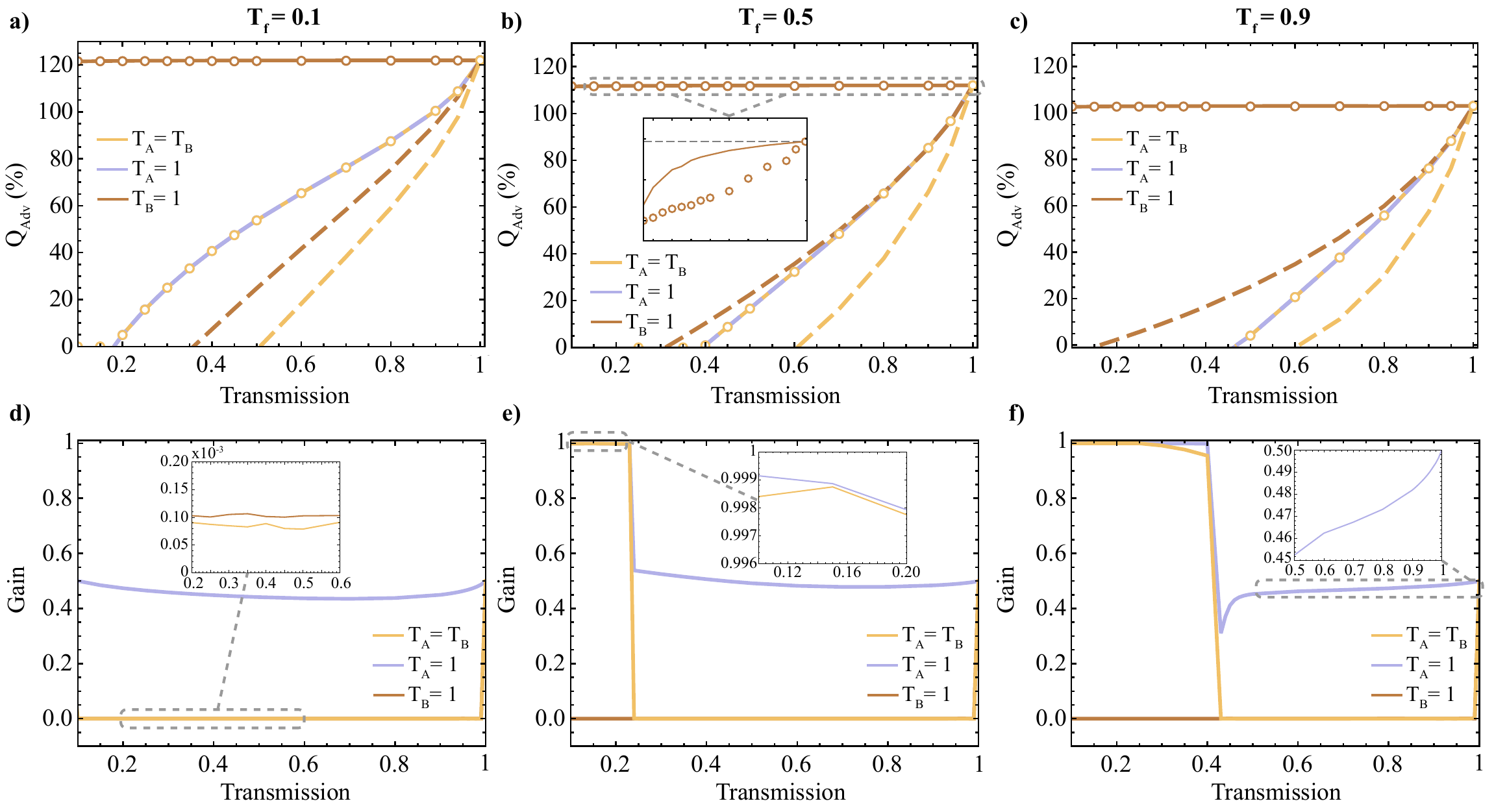}
\caption{\label{fig:figure8}Percentage improvement in the quantum advantage for superdense coding using the baseline, NLA/NA, and POVM schemes under fixed transmission loss when Alice transmits her encoded qubit to Bob. Orange lines correspond to the case where Charlie distributes qubits through symmetric pure-loss channels with equal transmissions $T_A = T_B$. The blue line represents scenarios where only Bob’s qubit experiences loss ($T_A = 1$, $T_B$ varies along the x-axis), while the brown line corresponds to loss on Alice’s qubit ($T_B = 1$, $T_A$ varies along the x-axis). Dashed lines indicate the quantum advantage of the baseline scheme over a single-qubit transmission, expressed as percentage improvement. Solid lines show the enhancement achieved with the NLA/NA circuit, and circular markers denote the improvements obtained using optimised POVMs. Panels (a--c) illustrate performance for different fixed channel transmissions when Alice sends her encoded qubit to Bob: \textbf{(a)} $T_f = 0.1$, \textbf{(b)} $T_f = 0.5$, and \textbf{(c)} $T_f = 0.9$. In panel (b), the horizontal grey dashed line in the subfigure represents the baseline quantum advantage at $T_f = 0.5$ with $T_A = T_B = 1$, which both the NLA/NA and POVM schemes converge towards. Panels (d--f) show the optimised NLA/NA transmittivities in the superdense coding protocol for different values of $T_f$. \textbf{(d)} $T_f = 0.1$: Although the transmittivities appear to approach $g_A = 0$, they remain finite, on the order of $10^{-3}$, as shown in the inset. \textbf{(e)} $T_f = 0.5$: Transmittivities near $g_A = 1$ remain slightly below unity, while those near zero are still non-zero, as indicated in the inset. \textbf{(f)} $T_f = 0.9$: Transmittivities near $g_A = 1$ are below unity and those near zero remain non-zero. For $g_A > 0.5$, Alice applies NA, and for $g_A < 0.5$, she applies NLA.
}
\end{figure*}
\begin{widetext}
\begin{equation}
\label{eq:NLA_superdense_matrix}
\rho_{AB}(p,\!g_A)=\frac{1}{2}
\begin{pmatrix}g_A\big(\!(1\!-\!p)(1\!-\!T_A)+p(1\!-\!T_B)\big) & 0 & 0 & 0 \\
0 & g_A\;p\;T_B & \sqrt{p(1-p)g_A(1-g_A)T_AT_B} & 0 \\
0 & \sqrt{p(1-p)g_A(1-g_A)T_AT_B} & (1-g_A)(1-p)T_A & 0 \\
0 & 0 & 0 & 0
\end{pmatrix}\!,
\end{equation}    
\end{widetext}
where $g_A$ denotes the beamsplitter transmittivity in the circuit of Fig.~\ref{fig:figure2}, which determines whether the circuit functions as an NLA or an NA. The success probability of the NLA/NA operation is given by $P_{\text{succ}} = 2\,\mathrm{Tr}[\rho_{AB}(p,g_A)]$, and the corresponding normalised state is 
\begin{equation}
    \label{eq:normalised_superdense_state_NLA}
    \Tilde{\rho}_{AB}(p,g_A) = \frac{\rho_{AB}(p,g_A)}{\mathrm{Tr}[\rho_{AB}(p,g_A)]}.
\end{equation}
Similar to the POVM optimisation, once Alice applies the NLA/NA, she encodes her classical bits using the corresponding Pauli operation and transmits the encoded qubit to Bob through a lossy channel. After this transmission, the two-qubit state becomes
\begin{align}
    \label{eq:rhoab_before_measurement_superdense_NLA_NA}
    \Tilde{\rho}_{AB}(p,g_A,\sigma_k,T_f)\!=&(\text{BS}(T_f)\otimes\mathbb{I}_2)(\sigma_k\otimes\mathbb{I}_2)\Tilde{\rho}_{AB}(p,g_A)\nonumber \\
    &(\sigma_k\otimes\mathbb{I}_2)^\dagger(\text{BS}(T_f)\otimes\mathbb{I}_2)^\dagger.
\end{align}
The cost function is again defined as the difference between the maximum mutual information shared by Alice and Bob and the classical channel capacity and expressed as
\begin{align}
    \label{eq:Qad_NLA_superdense}
    Q_{\text{NLA/NA}}(T_f)=&\!\max_{p,p_k,g_A}\bigg[S\bigg(\sum_{k=1}^4p_k\Tilde{\rho}_{AB}(p,g_A,\sigma_k,T_f)\bigg)\nonumber \\
    &-\sum_{k=1}^4p_kS\big(\Tilde{\rho}_{AB}(p,g_A,\sigma_k,T_f)\big)\bigg]-C(T_f),
\end{align}
where $g_A \in [0,1]$, subject to the additional requirement that the success probability satisfies $P_{\text{succ}} \geq 10^{-4}$, together with the constraints on $p$ and $p_k$ specified in Eq.~\eqref{eq:superdense_povm_constraint}.

\subsection{\label{eq:comparison_of_results_superdense_coding}Comparison of Superdense Coding Schemes}
Figure~\ref{fig:figure8} presents the percentage improvement in quantum advantage compared with the classical channel capacity. The figure of merit is defined as  
\begin{equation}
    \label{eq:percentage_improvement}
    Q_{\text{Adv}}(T_f) = \frac{Q_{\text{scheme}}(T_f) - C(T_f)}{C(T_f)} \times 100,
\end{equation}
where $Q_{\text{scheme}}(T_f)$ denotes the quantum advantage obtained under a given scheme (NA, NLA, or POVM).

The improvements are consistent across all transmission values shown in panels (a--c) of Fig.~\ref{fig:figure8}. In the case where Bob’s qubit is lossless and all loss occurs on Alice’s side (shown in brown), both the NLA/NA and POVM schemes saturate to the quantum advantage obtained at $T_A = T_B = 1$, regardless of the value of $T_A$. This convergence is most evident in panel (b), where the horizontal dark grey dashed line ($Q_{\text{Adv}} = 112\%$) in the subfigure indicates the baseline quantum advantage at $T_f = 0.5$ with $T_A = T_B = 1$, which both schemes approach. This behaviour is intuitive: since all loss is confined to Alice’s qubit before the encoding stage, her optimised POVM or the NLA/NA circuit can effectively correct for it locally before applying the encoding. Although the NLA/NA circuit achieves slightly higher improvements than the POVM optimisation, the difference is small. Moreover, as neither optimisation problem is convex, the reported results should be regarded as locally optimal rather than guaranteed global solutions.

When loss is present only on Bob’s qubit and Alice’s qubit is transmitted without loss ($T_A = 1$), all results coincide. Across panels (a--c) in Fig.~\ref{fig:figure8}, the solid blue line, dashed blue line, and blue circles correspond to the NLA/NA circuit, the baseline scheme, and the optimised POVM, respectively, and are found to overlap. Intuitively, this behaviour is expected since Alice’s POVM or the NLA/NA scheme cannot compensate for losses on Bob’s qubit. A more effective strategy would be to apply the correction at Bob’s station before he measures both qubits; however, once the success probability of this operation is taken into account, the mutual information between Alice and Bob would decrease, resulting in overall performance below the channel capacity. Therefore, while such a correction is beneficial in the teleportation scheme, where the figure of merit is the improvement in fidelity, it does not provide an advantage in superdense coding, since the overall improvement is penalised by the success probability when the process is applied in Bob's station.

However, when both Alice’s and Bob’s qubits experience loss ($T_A = T_B$) (shown in orange in Fig.~\ref{fig:figure8}), both the NLA/NA and POVM approaches outperform the baseline across all values of $T_f$. For instance, the solid orange line (NLA/NA) and the orange circles (POVM) coincide with the case of loss only on Bob’s qubit ($T_A = 1$), indicating that both processes are able to correct for losses on Alice’s qubit but cannot mitigate loss on Bob’s qubit. This outcome is expected, as the operation is applied solely on Alice’s side. Notably, when $T_A = T_B = 0.5$, the quantum advantage vanishes for $T_f = 0.1$, implying that superdense coding performs no better than the channel capacity, marking the break-even point. With a corrective process applied, however, such as POVM or NLA/NA, the break-even shifts to $T_A = T_B \approx 0.18$, enabling superdense coding to tolerate significantly greater loss. As the transmission $T_f$ between Alice and Bob increases, the shift in the break-even point becomes smaller. For instance, at $T_f = 0.5$ the break-even occurs at $T_A = T_B \approx 0.40$, while at $T_f = 0.9$ it shifts further to $\approx 0.45$. This behaviour is intuitive: when losses are low, even sending a single qubit (the classical case) can perform reasonably well. By contrast, at lower values of $T_f$ the impact of loss is more severe, and the combination of entanglement with corrective processes such as POVM or NLA/NA provides a clear advantage.  

Figure~\ref{fig:figure8}(d)--(f) presents the NLA/NA transmittivities applied by Alice in the superdense coding protocol, prior to her encoding operation. For $T_f = 0.1$, Alice consistently employs the NLA across all regimes, as indicated by transmittivity values below $g_A < 0.5$ for $T_A = 1$, $T_B = 1$, and $T_A = T_B$. The transmittivities for the $T_A = T_B$ and $T_B = 1$ regimes coincide, since Alice is the only party performing the NLA and can compensate only for losses on her side. In the case of $T_A = 1$, her use of the NLA is likewise intuitive, as the losses occur entirely on Bob’s side, and amplification serves to strengthen the shared entanglement between them. In panels (e) and (f), Alice’s strategy alternates between applying the NLA and NA depending on the transmission regime. For instance, at $T_f = 0.5$ and when $T_A = T_B \leq 0.23$, she initially applies the NA and then switches to the NLA. Similarly, for $T_B \leq 0.45$, Alice continues to apply the NA before transitioning to the NLA at higher transmission values. A comparable trend is observed in panel (f) for $T_f = 0.9$, where Alice applies the NA up to $T \leq 0.4$ for both $T_A = T_B$ and $T_A = 1$. Regardless of the value of $T_f$, however, when $T_B = 1$, Alice consistently applies the NLA across all transmissions. Overall, this indicates that Alice’s strategy depends not only on which of Charlie’s qubits experienced loss, but also on the additional losses her own encoded state incurs when transmitted to Bob.


\section{\label{sec:conclusion}Conclusion}
In this work, we investigated how \ozlem{SR-DV} teleportation and superdense coding can be enhanced in the presence of lossy channels by employing either general POVMs or the physically realisable processes of NA and NLA. By formulating the optimisation tasks for each protocol, we showed that the optimal POVMs reduce effectively to NA or NLA operations, depending on the transmission regime. This demonstrates that the simplest measurement strategies implementable in the laboratory can already capture the essential performance gains.

For \ozlem{SR-DV} teleportation, both the NLA/NA circuit and the optimised POVM enhance the average fidelity relative to the baseline scheme, achieving improvements of up to $75\%$ and $78\%$, respectively, in lossy channels, while maintaining feasible success probabilities. For superdense coding, NLA/NA and POVM strategies similarly enhance the quantum advantage over the classical channel capacity, with gains exceeding $100\%$ in certain regimes, thereby extending the tolerable range of losses and shifting the break-even points. Although the optimised POVMs offer marginal flexibility, the NLA/NA circuits provide nearly identical performance with direct experimental implementability.

\ozlem{Our results also indicate that SR-DV teleportation with satellite-distributed entanglement is realistically feasible. Even under substantial channel loss, Alice and Bob can restore the teleportation fidelity to values close to unity by locally applying a simple NLA or NA operations to the shared state they receive from the satellite, suggesting that high-quality state transfer is achievable in practical free-space links. By contrast, for SR-DV superdense coding, improvements over the classical capacity only arise when the satellite-ground transmissivity is above a certain threshold. This implies that, although superdense coding remains viable in free-space channels with moderate loss, teleportation is likely the more robust and broadly applicable primitive in realistic satellite-assisted quantum communication settings.}

We note that the optimisation landscapes in this work are likely to be non-convex, and the reported strategies correspond to locally optimal solutions. Nevertheless, the consistency of the improvements across both protocols highlights the robustness of NA and NLA as practical tools for mitigating losses in \ozlem{SR-DV} quantum communication. Given their simplicity and experimental accessibility, these operations provide a promising route for near-term implementations of loss-resilient \ozlem{SR-DV} teleportation and superdense coding, and more broadly, for strengthening the reliability of future quantum networks where teleportation plays a central role. 

\vspace{-1.5mm}
In this work, we focused on the pure-loss channel where photon loss is the only effect. A natural extension would be to consider more general channels, such as thermal-loss channels, where decoherence also plays a role. In such cases, NLA/NA circuits are still likely to provide an advantage over baseline schemes. Another promising direction would be to investigate the use of NLA/NA operations in \ozlem{SR-DV-QKD} repeater settings, where the circuit is applied to purify states prior to teleportation and subsequent entanglement distribution between parties for key extraction. This would make it possible to examine how the success probability of the process impacts the achievable secret key rates.

\section{\label{sec:discussion}Acknowledgments}
The Australian Government supported this research through the Australian Research Council’s Linkage Projects funding scheme (Project No. LP200100601). The views expressed herein are those of the authors and are not necessarily those of the Australian Government or the Australian Research Council.

\vspace{-1mm}
\appendix
\section{\label{appendix:results_with_phi_plus}Results Using an Alternative Distributed Bell State}
\begin{figure*}[t]
\includegraphics[scale=0.43]{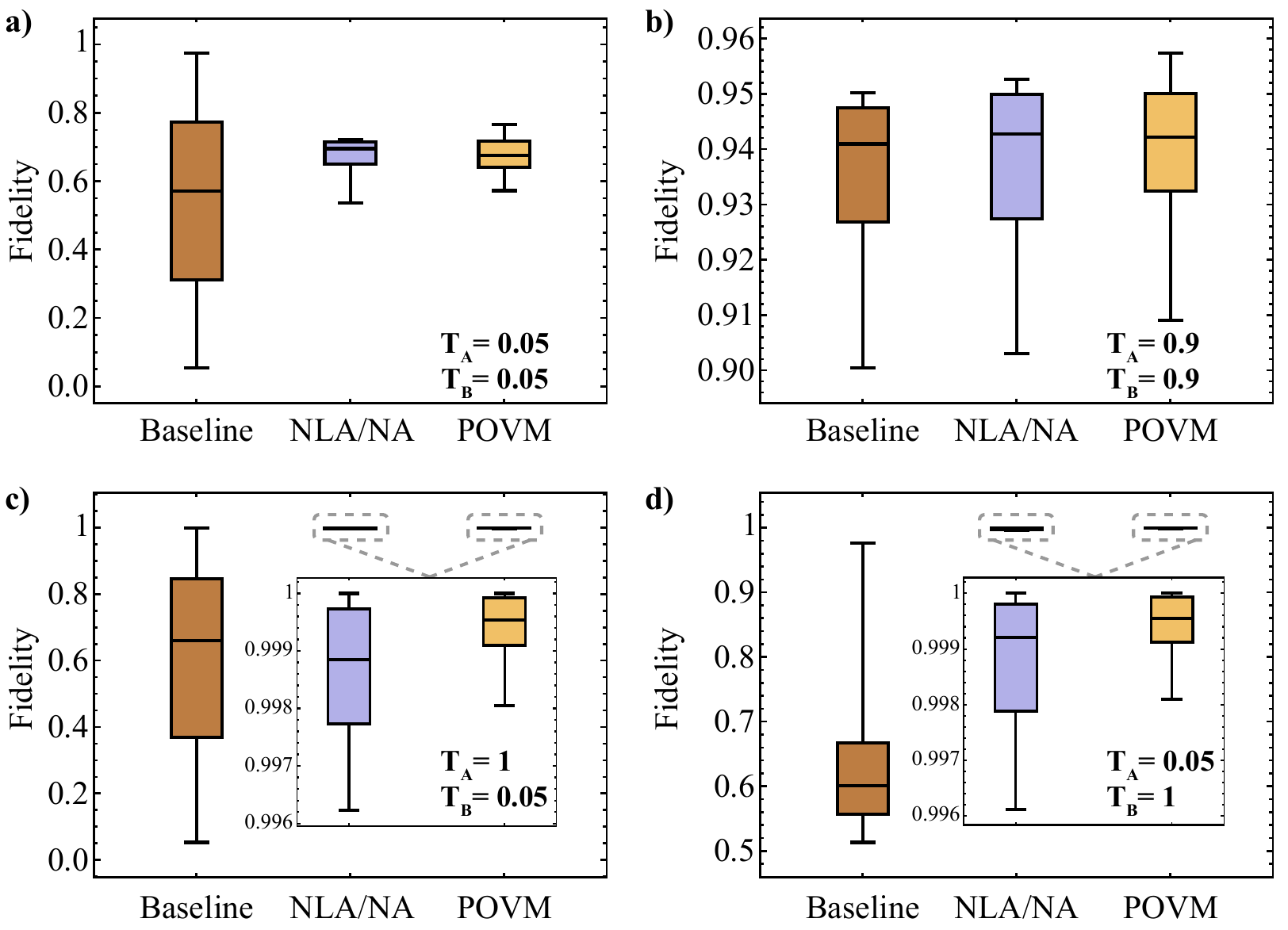}
\caption{\label{fig:figure9}Teleportation fidelities for $N = 300$ input states when the distributed entangled state is $\ket{\phi^+}$. Brown boxes show the baseline performance without any additional operation. Blue boxes correspond to teleportation where Alice and Bob apply NLA or NA, with the circuit transmittivities optimised to maximise the average fidelity. Orange boxes correspond to teleportation where Alice and Bob implement a POVM optimised for the best average fidelity. \textbf{(a)} Charlie distributes qubits to Alice and Bob with $T_A = T_B = 0.05$.  
\textbf{(b)} Charlie distributes qubits to Alice and Bob with $T_A = T_B = 0.9$. \textbf{(c)} Alice receives her qubit without loss ($T_A = 1$), while Bob’s qubit experiences a transmission of $T_B = 0.05$. \textbf{(d)} Bob receives his qubit without loss ($T_B = 1$), while Alice’s qubit experiences a transmission of $T_A = 0.05$.}
\end{figure*}

Throughout this paper, we assume that Charlie prepares the $\ket{\psi^+}$ Bell state and distributes one qubit each to Alice and Bob. To extend this analysis, let us now consider the case where Charlie instead prepares the following Bell state
\begin{equation}
    \label{eq:bell_state_phip}
    \ket{\phi_C}(p)=\sqrt{p}\ket{00}+\sqrt{1-p}\ket{11},
\end{equation}
where Charlie sets $p = 0.5$ for the teleportation schemes, while in the superdense coding protocol, the value of $p \in [0,1]$ is chosen to optimise the mutual information between Alice and Bob. Charlie then sends one qubit to Alice and the other to Bob through pure-loss channels with transmittivities $T_A$ and $T_B$, corresponding to their respective modes. The resulting shared state between Alice and Bob is therefore given by
\begin{widetext}
\begin{equation}
\label{eq:lossy_entangled_state_superdense_phip}
\rho_{AB}(p)=
\begin{pmatrix}p+(1-p)(1-T_A)(1-T_B) & 0 & 0 & \sqrt{p(1-p) T_A T_B} \\
0 & (1-p)(1-T_A)T_B & 0 & 0 \\
0 & 0 & (1-p)T_A(1-T_B) & 0 \\
\sqrt{p(1-p) T_A T_B} & 0 & 0 & (1-p)T_AT_B
\end{pmatrix}\!.
\end{equation}    
\end{widetext}
\subsection{\label{sec:results_with_phi_plus_teleportation}Results of the Teleportation Schemes}
In the teleportation scheme, we first consider the case where Alice and Bob apply a POVM to the state given in Eq.~\eqref{eq:lossy_entangled_state_superdense_phip} with $p = 0.5$. All other conditions remain identical to those in the main text and follow Eqs.~\eqref{eq:fidelity_maximisation}, ~\eqref{eq:povm_constraints}, and ~\eqref{eq:povm_prob_constraint}. However, the Bell-state projector $\Pi_{\psi^+}$ is replaced by $\Pi_{\phi^+} = \ket{\phi^+}\!\bra{\phi^+}$, where $\ket{\phi^+} = (\ket{00} + \ket{11})/\sqrt{2}$. Consequently, the corresponding success probability $P_{\psi^+}$ is replaced with $P_{\phi^+}$, associated with the Bell-state projection $\Pi_{\phi^+}$.

When Alice and Bob instead apply the NLA/NA circuit, the unnormalised shared state between them is given by
\begin{equation}
    \label{eq:general_state_AB_phip}
    \Tilde{\rho}_{AB}=\begin{pmatrix}
        a & 0 & 0 & b \\
        0 & c & 0 & 0 \\
        0 & 0 & d & 0 \\
       b & 0 & 0 & e   
    \end{pmatrix},
\end{equation}
with the matrix entries
\begin{equation}
\label{eq:general_state_entry_1_phip}
a=0.125g_{B}(1-g_{A})\big(1+(1-T_A)(1-T_B)\big),  
\end{equation}
\begin{equation}
    \label{eq:general_state_entry_2_phip}
    b=0.125\sqrt{g_A g_B(1-g_A)(1-g_B)T_AT_B},
\end{equation}
\begin{equation}
    \label{eq:general_state_entry_3_phip}
    c = 0.125(1-g_A)(1-g_B)(1-T_A)T_B,
\end{equation}
\begin{equation}
    \label{eq:general_state_entry_4_phip}
    d = 0.125g_Ag_BT_A (1-T_B),
\end{equation}
\begin{equation}
    \label{eq:general_state_entry_5_phip}
    e = 0.125g_A(1-g_B)T_A T_B.
\end{equation}
The success probability of the circuit is given as $P_{\text{succ}}=2\text{Tr}[\Tilde{\rho}_{AB}]$ and the normalised state becomes $\Tilde{\rho}_{AB}^n=\Tilde{\rho}_{AB}/\text{Tr}[\Tilde{\rho}_{AB}]$. In the optimisation, the cost function remains the same as in Eq.~\eqref{eq:fidelity_maximisation_nla_na}, with the constraints $0 \leq g_A \leq 1$ and $0 \leq g_B \leq 1$. The only modification is that the projection onto $\ket{\psi^+}$ is replaced by $\ket{\phi^+}$, and the corresponding success probability $P_{\psi^+}$ is replaced with $P_{\phi^+}$, associated with the Bell-state projection $\Pi_{\phi^+}$.

Figure~\ref{fig:figure9} illustrates the improvement in average \ozlem{SR-DV} teleportation fidelities achieved by the optimised NLA/NA and POVM operations, evaluated over $N = 300$ input states when the distributed entangled state is $\ket{\phi^+}$. The enhancements are less pronounced than those in Fig.~\ref{fig:figure4} for the symmetric-loss cases shown in Figs.~\ref{fig:figure9}(a) and (b). For instance, at $T_A = T_B = 0.05$, the median fidelities increase to $F = 0.97$ and $F = 0.98$ for the optimised NLA/NA circuits and POVMs, respectively, in Fig.~\ref{fig:figure4}(a) while the improvement achieved in Fig.~\ref{fig:figure9}(a) is $F = 0.70$ and $F = 0.68$ from a baseline median fidelity of $F = 0.58$. While the maximum fidelities remain unchanged in Fig.~\ref{fig:figure4}(a), they show a reduction in Fig.~\ref{fig:figure9}(a). However, the minimum fidelities increase, and both the NLA/NA and POVM optimisations yield a narrower and more stable fidelity distribution. For $T_A = T_B = 0.9$, the improvement in Fig.~\ref{fig:figure9}(b) is marginal compared to Fig.~\ref{fig:figure4}(b), as the median fidelities remain at the baseline value of $F = 0.94$ for both procedures, whereas in Fig.~\ref{fig:figure4}(b) the achieved median fidelity is $F = 0.9999$. In this high-transmission regime, the losses are minimal, and neither optimisation seem to lead to a noticeable benefit. However, when the losses are asymmetric, as shown in Figs.~\ref{fig:figure9}(c) and (d), the results are comparable to those in Figs.~\ref{fig:figure4}(c) and (d), achieving nearly identical fidelity values.
\begin{figure*}[t]
\includegraphics[scale=0.44]{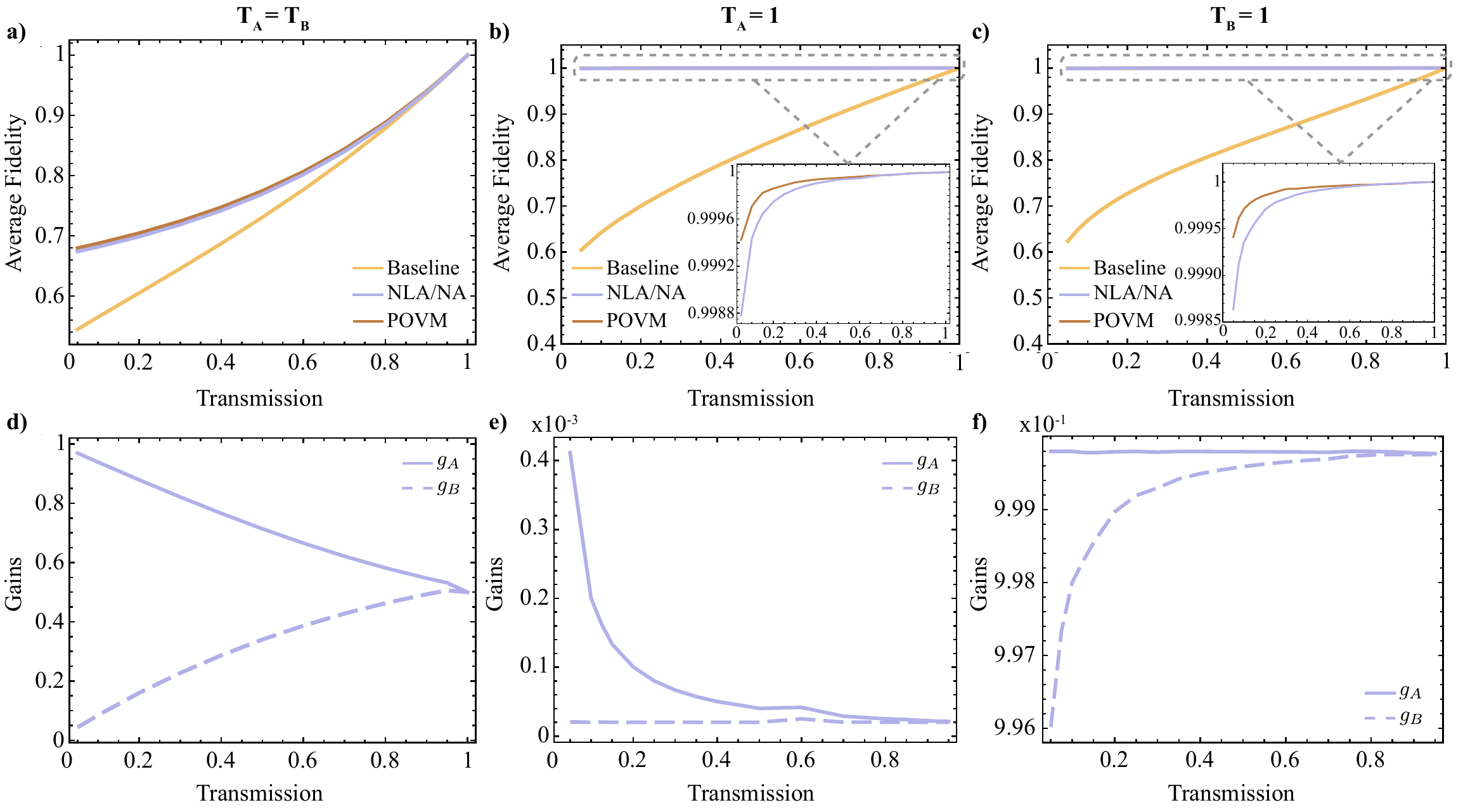}
\caption{\label{fig:figure10}Comparison of baseline, NLA/NA, and POVM teleportation across different transmission regimes when Charlie distributes the state $\ket{\phi^+}$. The results are shown in terms of average fidelities and optimised transmittivities. \textbf{(a)} Average fidelities for the three cases: baseline teleportation (orange), teleportation with NLA/NA using optimised transmittivities (blue), and teleportation with optimised POVMs (brown). In this panel, Charlie distributes the qubits to Alice and Bob through pure-loss channels with equal transmissions $T_A = T_B$. Panels \textbf{(b)} and \textbf{(c)} present the fidelities under asymmetric transmissions. In panel \textbf{(b)}, Alice’s qubit is transmitted without loss ($T_A = 1$), while Bob’s qubit has transmission $T_B$ varying from $0$ to $1$. Conversely, in panel \textbf{(c)}, Bob’s qubit is transmitted without loss ($T_B = 1$), while Alice’s qubit has transmission $T_A$. Panels \textbf{(d)}, \textbf{(e)}, and \textbf{(f)} show the optimised transmittivities for NLA/NA teleportation corresponding to the regimes $T_A = T_B$, $T_A = 1$, and $T_B = 1$, respectively.
}
\end{figure*}

Figure~\ref{fig:figure10}(a) further illustrates that in the symmetric-loss regime, Alice and Bob can improve the average fidelities but are unable to recover values close to unity, unlike in Fig.~\ref{fig:figure5}(a). At $T_A = T_B = 0.05$, the baseline average fidelity is $\bar{F} \approx 0.54$, which increases to $\bar{F} \approx 0.68$ with either optimisation method, an improvement of approximately $26\%$, as shown in Fig.~\ref{fig:figure10}(a). Under the same loss conditions, the corresponding improvements in Fig.~\ref{fig:figure5}(a) are $75\%$ and $78\%$ for the optimised POVM and NLA/NA schemes, respectively. However, for the asymmetric-loss regimes, Figs.~\ref{fig:figure10}(b) and (c) show improvements comparable to those in Figs.~\ref{fig:figure5}(b) and (c). This suggests that when Charlie distributes a single copy of $\ket{\phi^+}$ instead of $\ket{\psi^+}$, Alice and Bob are unable to recover the entangled state in the symmetric-loss case. In contrast, under asymmetric-loss conditions, they are able to successfully recover the shared entanglement. This behaviour is directly related to the fidelity of the teleported state, as a less degraded shared entangled state yields higher teleportation fidelity.

Figures~\ref{fig:figure10}(d)--(f) present the optimised NLA/NA transmittivities that maximise the average fidelities across different transmission regimes. In the symmetric case, Alice’s transmittivities are consistently above $g_A \geq 0.5$ while Bob’s are below $g_B \leq 0.5$, indicating that Alice applies NA and Bob applies NLA. At $T_A = T_B = 1$, both transmittivities converge to $g_A = g_B = 0.5$, corresponding to no operation since the system is lossless. This behaviour is consistent with the strategies observed in Fig.~\ref{fig:figure5}(d). When Alice’s qubit is transmitted without loss ($T_A = 1$) and Bob’s qubit undergoes partial transmission, the optimal strategy is for both parties to apply NLA, with transmittivities below 0.5. In contrast, when Bob’s qubit is lossless ($T_B = 1$) and Alice’s qubit experiences transmission loss, both parties apply NA, with transmittivities above 0.5. However, this behaviour differs from that observed in Figs.~\ref{fig:figure5}(e) and (f), where Alice and Bob consistently adopt opposite strategies. In both cases, Alice applies the NA, while Bob applies the NLA.

\subsection{\label{sec:results_with_phi_plus_superdense}Results of the Superdense Coding}
In the superdense coding scheme, we first consider the case where Alice applies a POVM to her received qubit before performing the encoding operation. In this scenario, all equations from~\eqref{eq:rhoab_superdense}--\eqref{eq:superdense_povm_constraint} remain valid for the POVM optimisation when Charlie distributes the $\ket{\phi^+}$ state to Alice and Bob, except that $\rho_{AB}(p)$ in Eq.~\eqref{eq:rhoab_superdense} is now replaced with the definition given in Eq.~\eqref{eq:lossy_entangled_state_superdense_phip}.

For the NLA/NA operation, all the other equations remain unchanged, except that Eq.~\eqref{eq:NLA_superdense_matrix} is replaced with the new matrix, expressed as
\begin{widetext}
\begin{equation}
\label{eq:NLA_superdense_matrix_phip}
\rho_{AB}(p,\!g_A)\!=\!\frac{1}{2}\!\!
\begin{pmatrix}g_A\big(\!(1\!-\!p)(1\!-\!T_A)(1\!-\!T_B)\!-\!p\big) & 0 & 0 & \sqrt{g_A(1\!-\!g_A)p(1\!-\!p) T_A T_B} \\
0 & \!\!\!\!\!g_A(p\!-\!1)(1\!-\!T_A)T_B & 0 & 0 \\
0 & 0 & \!\!\!\!(g_A\!-\!1)(1\!-\!p)T_A(1\!-\!T_B) & 0 \\
\sqrt{g_A(1\!-\!g_A)p(1\!-\!p) T_A T_B} & 0 & 0 & \!\!\!\!\!\!\!\!\!\!\!\!\!(g_A\!-\!1)(1\!-\!p)T_AT_B
\end{pmatrix}\!.
\end{equation}
\end{widetext}

\begin{figure*}[t]
\includegraphics[scale=0.46]{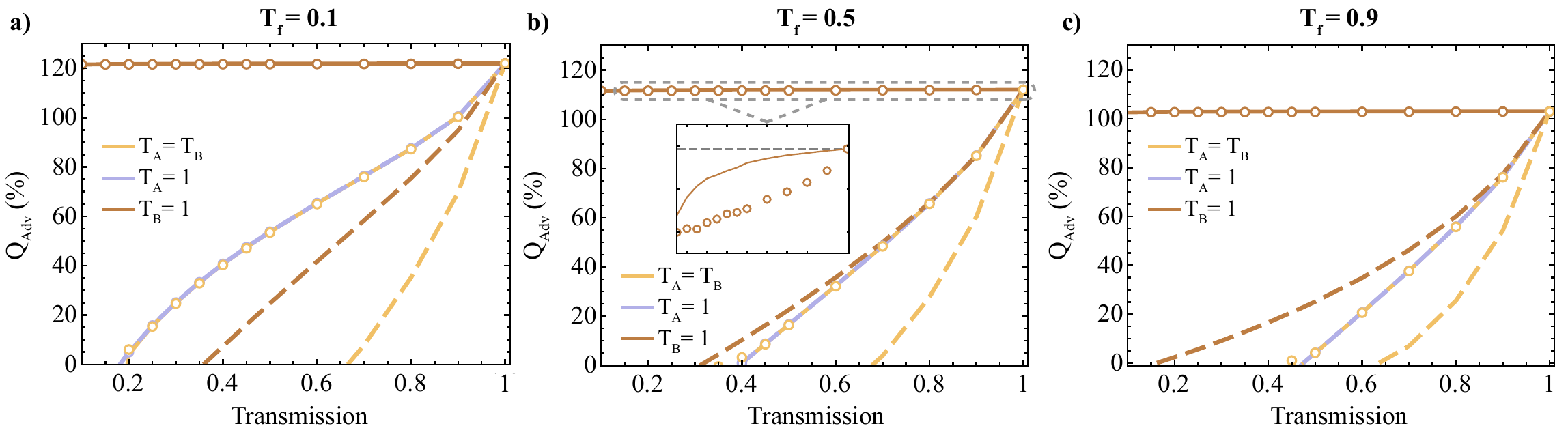}
\caption{\label{fig:figure11}Percentage improvement in the quantum advantage for superdense coding using the baseline, NLA/NA, and POVM schemes when the distributed entangled state is $\ket{\phi^+}$, under fixed transmission loss as Alice transmits her encoded qubit to Bob. Orange lines correspond to the case where Charlie distributes qubits through symmetric pure-loss channels with equal transmissions $T_A = T_B$. The blue line represents scenarios where only Bob’s qubit experiences loss ($T_A = 1$, $T_B$ varies along the x-axis), while the brown line corresponds to loss on Alice’s qubit ($T_B = 1$, $T_A$ varies along the x-axis). Dashed lines indicate the quantum advantage of the baseline scheme over a single-qubit transmission, expressed as percentage improvement. Solid lines show the enhancement achieved with the NLA/NA circuit, and circular markers denote the improvements obtained using optimised POVMs. Panels (a--c) illustrate performance for different fixed channel transmissions when Alice sends her encoded qubit to Bob: \textbf{(a)} $T_f = 0.1$, \textbf{(b)} $T_f = 0.5$, and \textbf{(c)} $T_f = 0.9$. In panel (b), the horizontal grey dashed line in the subfigure represents the baseline quantum advantage at $T_f = 0.5$ with $T_A = T_B = 1$, which both the NLA/NA and POVM schemes converge towards.
}
\end{figure*}

Figure~\ref{fig:figure11}(a--c) presents the results of the superdense coding protocol using both NLA/NA and optimised POVMs when Charlie distributes the $\ket{\phi^+}$ state across different loss regimes. The overall trends are consistent with those observed in Fig.~\ref{fig:figure8}(a--c). For all values of $T_f$ shown in panels (a)--(c), when the loss occurs on Alice’s qubit and Bob’s channel remains lossless ($T_B = 1$), the maximum quantum advantage matches that obtained with the $\ket{\psi^+}$ state in Fig.~\ref{fig:figure8}(a--c). This behaviour is intuitive, as losses on Alice’s side can be effectively compensated by her NLA/NA circuit or optimised POVMs. However, when $T_A = T_B$, the baseline curves shift slightly to the right along the x-axis. This shift indicates that the break-even points occur at higher transmission values, meaning that better channel quality is required to achieve a quantum advantage in Fig.~\ref{fig:figure11}(a)--(c) compared with Fig.~\ref{fig:figure8}(a)--(c) for the baseline superdense coding scheme. For instance, when $T_f = 0.1$ in Fig.~\ref{fig:figure11}(a), the break-even point occurs at $T_A = T_B = 0.6$, whereas in Fig.~\ref{fig:figure8}(a) it appears at $T_A = T_B = 0.5$. After applying either optimisation scheme, the curves converge to the $T_A = 1$ line, matching the break-even point of $0.18$ observed in Fig.~\ref{fig:figure8}(a). Overall, the similar results observed in the superdense coding scheme can be explained by the fact that only Alice performs an active operation. Therefore, regardless of which Bell state Charlie distributes, the best she can do is to counteract the loss affecting her own qubit, unlike the teleportation schemes where both parties can apply local operations assisted by classical communication to recover the shared entanglement.

\section{\label{appendix:proof_symmetric_loss}Recoverability of Bell States Under Symmetric Loss}
In this section, we explicitly demonstrate that when the qubits of a Bell state experience symmetric imperfections during distribution from Charlie to Alice and Bob, not all Bell states remain equally useful for quantum communication. Our analysis is restricted to the single-rail setting, where only a single copy of the Bell state is shared between Alice and Bob and is consistent with the general two-qubit local-filtering/quasidistillation framework developed in Refs.~\cite{horodecki1999general,verstraete2001local}. In particular, we show that the state $\ket{\psi^+} = ( \ket{01} + \ket{10} ) / \sqrt{2}$ can be approximately recovered through local operations and classical communication~(LOCC) within a finite deviation from the ideal case, whereas the state $\ket{\phi^+} = ( \ket{00} + \ket{11} ) / \sqrt{2}$ cannot be recovered, even approximately, under symmetric conditions.

\subsection{\label{appendix:proof_psiplus}Proof of Recoverability of $\ket{\psi^+}$ up to a Threshold}
\textit{Lemma 1} (Approximate recoverability of $\ket{\psi^+}$ under symmetric loss). In the single-rail setting, where only one copy of the Bell state is distributed from Charlie to Alice and Bob, suppose each qubit undergoes an independent pure-loss channel with transmittivities $T_A, T_B \in [0,1]$. Then there exists an LOCC branch, implemented by a product Kraus operator $K = A \otimes B$, with non-zero success probability such that, when applied to the lossy $\ket{\psi^+}$ state, the conditional output state can be made arbitrarily close to $\ket{\psi^+}$.

More precisely, for any $\varepsilon > 0$ there exists a filter strength $\delta > 0$ for which 
\begin{itemize}
    \item the success probability scales as $P_s = O(\delta^2)$, and 
    \item the conditional fidelity satisfies $F \geq 1 - \varepsilon$.
\end{itemize}
Hence $\ket{\psi^+}$ is approximately recoverable under symmetric loss up to an arbitrarily small error, at the cost of a quadratic reduction in success probability.

\textit{Proof.} The shared state between Alice and Bob after Charlie distributes the qubits of $\ket{\psi^+}$ through the pure-loss channels is given in Eq.~\eqref{eq:lossy_bell_state}, which can also be rewritten as
\begin{align}
\rho_{AB} = &p_{00}\ketbra{00}{00} 
+ p_{01}\ketbra{01}{01} 
+ \gamma\ketbra{01}{10}+ \nonumber \\
&\gamma\ketbra{10}{01} 
+ p_{10}\ketbra{10}{10},
\end{align}
where
\begin{equation}
    p_{00}=0.5\big((1-T_A)+(1-T_B)\big),
\end{equation}
\begin{equation}
    p_{01}=0.5T_B,
\end{equation}
\begin{equation}
    p_{10}=0.5T_A,
\end{equation}
\begin{equation}
    \gamma=0.5\sqrt{T_AT_B}
\end{equation}

Let $A$ and $B$ denote single-qubit Kraus operators acting locally on Alice’s and Bob’s modes, respectively. The overall local filter is then represented by the product operator $K = A \otimes B$, which defines a single branch of a heralded LOCC protocol. Each local filter acts non–trace-preservingly on the single-rail qubit basis according to
\begin{equation}
\label{eq:operator_conditions}
A\ket{0} = a_0, \quad A\ket{1} = a_1, \qquad 
B\ket{0} = b_0, \quad B\ket{1} = b_1,
\end{equation}
where $a_0, a_1, b_0, b_1$ are (in general unnormalised) single-qubit vectors. 
We introduce a small scaling parameter $\delta > 0$, representing a perturbation close to zero up to a threshold, and fixed constants $\alpha, \beta > 0$ such that
\begin{equation}
\label{eq:vector_conditions}
a_0 = \delta\,\hat{a}_0, \quad b_0 = \delta\,\hat{b}_0, \qquad 
a_1 = \alpha\,\hat{a}_1, \quad b_1 = \beta\,\hat{b}_1,
\end{equation}
with $\hat{a}_i, \hat{b}_i$ unit vectors defining the filter directions. 
This parametrisation suppresses the vacuum component $\ket{00}$ by $\delta^2$ while leaving the one-photon subspace $\mathrm{span}\{\ket{01},\ket{10}\}$ scaled by $\delta$.

Applying the local filter $K$ to the lossy Bell state $\rho_{AB}$, the resulting components of $K\rho_{AB}K^{\dagger}$ can be written as
\begin{equation}
p_{00}K\ketbra{00}{00}K^{\dagger} = \delta^4p_{00}\ketbra{\Hat{a}_0\Hat{b}_0}{\Hat{a}_0\Hat{b}_0},
\end{equation}
\begin{equation}
p_{01}K\ketbra{01}{01}K^{\dagger} = \delta^2\beta^2p_{01}\ketbra{\Hat{a}_0\Hat{b}_1}{\Hat{a}_0\Hat{b}_1},
\end{equation}
\begin{equation}
\gamma K\ketbra{01}{10}K^{\dagger} = \delta^2\alpha \beta \gamma \ketbra{\Hat{a}_0\Hat{b}_1}{\Hat{a}_1\Hat{b}_0},
\end{equation}
\begin{equation}
\gamma K\ketbra{10}{01}K^{\dagger} = \delta^2\alpha \beta \gamma \ketbra{\Hat{a}_1\Hat{b}_0}{\Hat{a}_0\Hat{b}_1},
\end{equation}
\begin{equation}
p_{10}\ketbra{10}{10}K^{\dagger} = \delta^2\alpha^2 p_{10} \ketbra{\Hat{a}_1\Hat{b}_0}{\Hat{a}_0\Hat{b}_1}.
\end{equation}
The success probability of the this operation is given as $\text{Tr}(K\rho_{AB}K^{\dagger})$, which is simply expressed as
\begin{align}
    \label{eq:success_prob_LOCC_psip}
    P_{\mathrm{succ}}&=\mathrm{Tr}(K\rho_{AB}K^{\dagger})=\delta^4p_{00}+\delta^2\beta^2p_{01}+\delta^2\alpha^2 p_{10}, \nonumber \\
    P_{\mathrm{succ}}&\approx\delta^2
\end{align}
where $\alpha$, $\beta$, $p_{00}$, $p_{01}$, and $p_{10}$ are all much larger than $\delta$ (for example, taking $\delta = 10^{-4}$). In this case, the success probability is dominated by the $\delta^2$ term. In this case, the normalised post-selected state after the LOCC operation is given by $(K\rho_{AB}K^{\dagger})/P_{\mathrm{succ}}$, where each component of the state can be written as
\begin{equation}
\label{eq:suppresed_vacuum}
\frac{p_{00}K\ketbra{00}{00}K^{\dagger}}{P_{\mathrm{succ}}} = \delta^2p_{00}\ketbra{\Hat{a}_0\Hat{b}_0}{\Hat{a}_0\Hat{b}_0},
\end{equation}
\begin{equation}
\frac{p_{01}K\ketbra{01}{01}K^{\dagger}}{P_{\mathrm{succ}}} = \beta^2p_{01}\ketbra{\Hat{a}_0\Hat{b}_1}{\Hat{a}_0\Hat{b}_1},
\end{equation}
\begin{equation}
\frac{\gamma K\ketbra{01}{10}K^{\dagger}}{P_{\mathrm{succ}}} = \alpha \beta \gamma \ketbra{\Hat{a}_0\Hat{b}_1}{\Hat{a}_1\Hat{b}_0},
\end{equation}
\begin{equation}
\frac{\gamma K\ketbra{10}{01}K^{\dagger}}{P_{\mathrm{succ}}} = \alpha \beta \gamma \ketbra{\Hat{a}_1\Hat{b}_0}{\Hat{a}_0\Hat{b}_1},
\end{equation}
\begin{equation}
\frac{p_{10}\ketbra{10}{10}K^{\dagger}}{P_{\mathrm{succ}}} = \alpha^2 p_{10} \ketbra{\Hat{a}_1\Hat{b}_0}{\Hat{a}_0\Hat{b}_1}.
\end{equation}
Without loss of generality, we choose the filter directions aligned with the computational basis, i.e. set $\hat{a}_0=\ket{0}$, $\hat{a}_1=\ket{1}$, $\hat{b}_0=\ket{0}$, and $\hat{b}_1=\ket{1}$. With this identification we have $\ket{\hat{a}_0\hat{b}_0}\equiv\ket{00}$, $\ket{\hat{a}_0\hat{b}_1}\equiv\ket{01}$, and $\ket{\hat{a}_1\hat{b}_0}\equiv\ket{10}$, so the filtered components above are directly expressed in the computational basis. Then the overall state can be expressed as
\begin{align}
\label{eq:normalised_new_psi_plus}
    \Tilde{\rho}_{AB} = &\delta^2 p_{00}\ketbra{00}{00} 
+ \beta^2 p_{01}\ketbra{01}{01} 
+ \alpha \beta \gamma\ketbra{01}{10}+ \nonumber \\
&\alpha \beta \gamma \ketbra{10}{01} 
+ \alpha^2 p_{10}\ketbra{10}{10}.
\end{align}
Therefore, after normalisation, the vacuum component $\ketbra{00}{00}$ becomes smaller by a factor of $\delta^2$ as shown in Eq.~\eqref{eq:suppresed_vacuum} compared to the one-photon contributions, and can thus be neglected in the small-$\delta$ limit.

After applying the LOCCs, we choose the transmittivities so that the diagonal weights of the one-photon block are equal, i.e., the $\ketbra{01}{01}$ and $\ketbra{10}{10}$ terms carry the same coefficient, where we impose
\begin{align}
    \beta^2 p_{01} &= \alpha^2 p_{10}\nonumber \\
    \frac{\beta^2}{\alpha^2} &= \frac{p_{10}}{p_{01}}\nonumber \\
    \frac{\beta}{\alpha} &= \sqrt{\frac{p_{10}}{p_{01}}} \nonumber \\
    \frac{\beta}{\alpha} &= \sqrt{\frac{T_A}{T_B}}
\end{align}
In the symmetric loss regime ($T_A = T_B$), this condition ensures that both arms are equally weighted, implying $\beta=\alpha$. In this case, the overall state can be expressed as
\begin{align}
\label{eq:normalised_new_psi_plus_simpler}
    \Tilde{\rho}_{AB} = &\delta^2 p_{00}\ketbra{00}{00} 
+ 0.5\alpha^2 T_A\ketbra{01}{01} 
+ 0.5\alpha^2 T_A\ketbra{01}{10} \nonumber \\
&+ 0.5\alpha^2 T_A\ketbra{10}{01} 
+ 0.5\alpha^2 T_A\ketbra{10}{10},
\end{align}
which is simply equal to
\begin{align}
\label{eq:normalised_state_simplified}
     \Tilde{\rho}_{AB}&=\delta^2 p_{00}\ketbra{00}{00}+\alpha^2 T_A\ketbra{\psi^+}{\psi^+}\nonumber \\
     &=\lambda\ketbra{00}{00}+(1-\lambda)\ketbra{\psi^+}{\psi^+}.
\end{align}
As this state is normalised, we have $1 - \lambda = \delta^2 p_{00}$, where $\delta$ is a small but a non-zero parameter (e.g., $\delta = 10^{-4}$). Since $p_{00} \gg \delta$, the product $\delta^2 p_{00}$ remains dominated by $\delta^2$, and we can therefore approximate $1 - \lambda \approx \delta^2$. Then the fidelity of this state with respect to $\ket{\psi^+}$ is therefore given by
\begin{equation}
    F = \bra{\psi^+}\Tilde{\rho}_{AB}\ket{\psi^+} = 1 - \lambda \approx 1 - \delta^2,
\end{equation}
where the approximation holds since $\delta^2 p_{00} \approx \delta^2$ in the small-$\delta$ limit.

\subsection{\label{appendix:proof_phi_plus}Proof of Non-recoverability of $\ket{\phi^+}$}
\textit{Lemma 2} (Non-recoverability of $\ket{\phi^+}$ under symmetric loss). Under symmetric pure-loss ($T_A = T_B$) in the single-rail, single-copy regime, $\ket{\phi^+}$ cannot be recovered by any heralded LOCC operation of product form $K = A \otimes B$ with non-zero success probability, even approximately within any finite fidelity threshold.

\textit{Proof.} The shared state between Alice and Bob after Charlie distributes the qubits of $\ket{\phi^+}$ through the pure-loss channels is given in Eq.~\eqref{eq:lossy_entangled_state_superdense_phip}, which can also be rewritten as
\begin{align}
\rho_{AB} = &p_{00}\ketbra{00}{00} 
+ \gamma\ketbra{00}{11} 
+ p_{01}\ketbra{01}{01}+\nonumber \\
&p_{10}\ketbra{10}{10}+ \gamma\ketbra{11}{00}+p_{11}\ketbra{11}{11}
\end{align}
where
\begin{equation}
    p_{00}=0.5\big(1+(1-T_A)(1-T_B)\big),
\end{equation}
\begin{equation}
    p_{01}=0.5(1-T_A)T_B,
\end{equation}
\begin{equation}
    p_{10}=0.5T_A(1-T_B),
\end{equation}
\begin{equation}
    p_{11}=0.5T_AT_B,
\end{equation}
\begin{equation}
    \gamma=0.5\sqrt{T_AT_B}
\end{equation}

Let \textit{A} and \textit{B} denote single-qubit Kraus operators acting locally on Alice's and Bob's modes, separately. We follow the same notation as Appendix~\ref{appendix:proof_psiplus} for $A$ and $B$ where they follow Eq.~\eqref{eq:operator_conditions}. The individual components of the joint operator $K = A \otimes B$ acting on the shared state can therefore be expressed through its action on the computational basis
\begin{equation}
    K\ket{00}=a_0\otimes b_0,
\end{equation}
\begin{equation}
    K\ket{01}=a_0\otimes b_1
\end{equation}
\begin{equation}
    K\ket{10}=a_1\otimes b_0,
\end{equation}
\begin{equation}
    K\ket{11}=a_1\otimes b_1,
\end{equation}
where we require $\ket{01},\ket{10}$ to vanish to recover the $\ket{\phi^+}$. This implies that 
\begin{equation}
    \label{eq:conditions_for_fidelity}
    a_0\otimes b_1\approx0 \quad \mathrm{and} \quad a_1\otimes b_0\approx0,
\end{equation}
which in turn implies
\begin{equation}
    \label{eq:conditions_for_fidelity2}
    a_0 = 0 \lor b_1 = 0 \quad \mathrm{and} \quad a_1 = 0 \lor b_0 = 0.
\end{equation}
However, note that we wish to keep $\ket{00},\ket{11}$ components which also implies that 
\begin{equation}
    \label{eq:conditions_for_fidelity3}
    a_0\otimes b_0\neq0 \quad \mathrm{and} \quad a_1\otimes b_1\neq0,
\end{equation}
which in turn implies that all local components $a_0, a_1, b_0, b_1$  must be non-zero and therefore the cross-term constraints $a_0 \otimes b_1 = 0$ and $a_1 \otimes b_0 = 0$ cannot hold simultaneously. Even if we use a similar definition to Eq.~\ref{eq:vector_conditions}, where the vectors are close to zero but not exactly vanishing, we can make the following approximations
\begin{equation}
    \label{eq:approximate_conditions1}
    a_0\approx\delta\hat{a}_0 \;\;\text{and}\;\;  a_1\approx\delta\hat{a}_1 \Longrightarrow
    b_0 \approx \beta \hat{b}_0 \;\;\text{and} \;\; b_1 \approx \beta \hat{b}_1,
\end{equation}
\begin{equation}
    \label{eq:approximate_conditions2}
    a_0\approx\delta\hat{a}_0 \;\; \text{and} \;\; b_0\approx\delta\hat{b}_0 \Longrightarrow
    a_1 \approx\alpha\hat{a}_1 \;\;\text{and} \;\; b_1 \approx \beta \hat{b}_1,
\end{equation}
\begin{equation}
    \label{eq:approximate_conditions3}
    b_1\approx\delta\hat{b}_1 \;\;\text{and} \;\; a_1\approx\delta\hat{a}_1
    \Longrightarrow
    b_0 \approx\beta\hat{b}_0 \;\;\text{and} \;\; a_0 \approx \alpha \hat{a}_0,
\end{equation}
\begin{equation}
    \label{eq:approximate_conditions4}
    b_1\approx\delta\hat{b}_1 \;\; \text{and} \;\; b_0\approx\delta\hat{b}_0
    \Longrightarrow
    a_0 \approx\alpha\hat{a}_0 \;\;\text{and} \;\; a_1 \approx \alpha \hat{a}_1.
\end{equation}

Based on these approximations, we examine two representative cases, as the remaining two yield similar results. Using the relations in Eq.~\eqref{eq:approximate_conditions1}, the action of the Kraus operator on the computational basis states can be written as
\begin{equation}
    K\ket{00}=a_0\otimes b_0\approx\delta \ket{\hat{a}_0\hat{b}_0},
\end{equation}
\begin{equation}
    K\ket{01}=a_0\otimes b_1\approx\alpha \delta \ket{\hat{a}_0\hat{b}_1}
\end{equation}
\begin{equation}
    K\ket{10}=a_1\otimes b_0\approx\beta \delta \ket{\hat{a}_1\hat{b}_0},
\end{equation}
\begin{equation}
    K\ket{11}=a_1\otimes b_1\approx\alpha \delta \ket{\hat{a}_1\hat{b}_1}.
\end{equation}
We observe that the undesired components $\ket{01}$ and $\ket{10}$ are indeed suppressed. However, this suppression also affects $\ket{00}$ and $\ket{11}$, preventing recovery of the target state $\ket{\phi^+}$. We now examine Eq.~\eqref{eq:approximate_conditions2} to verify this behaviour.
\begin{equation}
    K\ket{00}=a_0\otimes b_0\approx\delta^2 \ket{\hat{a}_0\hat{b}_0},
\end{equation}
\begin{equation}
    K\ket{01}=a_0\otimes b_1\approx\beta \delta \ket{\hat{a}_0\hat{b}_1}
\end{equation}
\begin{equation}
    K\ket{10}=a_1\otimes b_0\approx\alpha \delta \ket{\hat{a}_1\hat{b}_0},
\end{equation}
\begin{equation}
    K\ket{11}=a_1\otimes b_1\approx\alpha \beta \ket{\hat{a}_1\hat{b}_1}.
\end{equation}
Although the $\ket{11}$ term remains unaffected, the $\ket{00}$ component is still suppressed, confirming that the target state $\ket{\phi^+}$ cannot be recovered even with a small threshold.

\section{\label{appendix:POVM_NLA_Matrices}Explicit Examples of Optimised Local POVMs for SR-DV Teleportation}
\begin{figure}[t]
\hspace*{-0.7cm}
\includegraphics[scale=0.63]{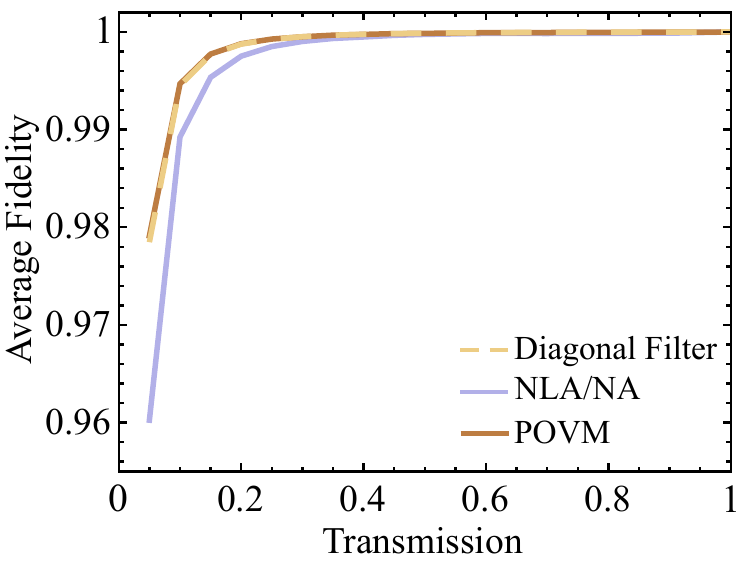}
\caption{\label{fig:figure_appendix1}Optimised average teleportation fidelity under symmetric transmission, $T_A = T_B$, for three filtering strategies: general POVMs (brown), general diagonal filters (yellow), and NLA/NA operations (blue). All optimisations were performed under the same success probability constraint, $P_{\mathrm{succ}} \ge 10^{-4}$. The general diagonal filters nearly coincide with the optimised POVMs and both show a modest advantage over NLA/NA at low transmission. At higher transmission, all three approaches become nearly identical.}
\end{figure}
In this section, we provide representative examples of the optimised local POVMs used in the SR-DV teleportation analysis of Fig.~\ref{fig:figure5}, and compare them with the corresponding optimised NLA/NA operations. For the symmetric-loss case, where Alice and Bob’s qubits experience equal transmission $T_A=T_B=0.1$, the optimised local operators $M_{A_1}$ and $M_{B_1}$ are given by
\begin{equation}
\label{eq:matrix_A_equal_loss}
M_{A_1}=\begin{pmatrix}
    0.0274-0.0072i & 0.4407+0.0614i \\
    -0.0122+0.0070i & 0.8889-0.1091i
\end{pmatrix},
\end{equation}
\begin{equation}
\label{eq:matrix_B_equal_loss}
M_{B_1}=\begin{pmatrix}
    0.0281+0.0011i & -0.4034-0.1874i \\
    0.0136-0.0031i & 0.8821+0.1551i
\end{pmatrix}.
\end{equation}
The corresponding POVM elements are then defined as
\begin{equation}
\label{eq:povm_elements}
E_{A_1}\approx E_{B_1}\approx\begin{pmatrix}
    0.0010 & 0 \\
    0 & 0.9999
\end{pmatrix}.
\end{equation}
In this case, the effect of the operators are nearly identical for Alice and Bob. The dominant action is a strong suppression of the $\ket{0}$ component, while the $\ket{1}$ component is left almost unchanged.

\ozlemREV{To clarify the origin of the slight advantage of the optimised POVMs over the NLA/NA filters, we also consider an intermediate optimisation over general local diagonal filters. This allows us to separate the effect of a more general diagonal vacuum-suppression filter from that of the full off-diagonal/unitary freedom present in a general Kraus operator. Specifically, we optimise local filters of the form
\begin{equation}
M_{A_1}^{\mathrm{diag}}=\begin{pmatrix}
a_0 & 0 \\
0 & a_1
\end{pmatrix},
\qquad
M_{B_1}^{\mathrm{diag}}=\begin{pmatrix}
b_0 & 0 \\
0 & b_1
\end{pmatrix},
\end{equation}
with $0\leq a_0,a_1,b_0,b_1\leq 1$, under the same success probability constraint
\begin{equation}
P_{\mathrm{succ}} \geq 10^{-4}.
\end{equation}
For the symmetric-loss case with $T_A=T_B=0.1$, the optimisation yields identical diagonal filters for Alice and Bob,
\begin{equation}
M_{A_1}^{\mathrm{diag}} = M_{B_1}^{\mathrm{diag}} =
\begin{pmatrix}
0.0316 & 0 \\
0 & 1.0000
\end{pmatrix},
\end{equation}
with corresponding POVM elements
\begin{equation}
E_{A_1}^{\mathrm{diag}} = E_{B_1}^{\mathrm{diag}} =
\begin{pmatrix}
0.0010 & 0 \\
0 & 1.0000
\end{pmatrix},
\end{equation}
which matches the optimised general POVM elements given in Eq.~\eqref{eq:povm_elements}.}

\ozlemREV{The close overlap between the optimised diagonal-filter and POVM curves in Fig.~\ref{fig:figure_appendix1} indicates that the main useful action of the optimised POVMs is already largely diagonal in the single-rail basis. In the symmetric-loss case, the optimised diagonal filters closely reproduce the performance of the full POVMs, while both show a modest advantage over the constrained NLA/NA filters in the low-transmission regime. This indicates that the small gap between the POVM and NLA/NA results does not primarily arise from off-diagonal or unitary freedom of the POVMs, but rather from the fact that the NLA/NA scheme explores only a restricted subset of the more general diagonal filter family.}

For comparison, the optimised NLA/NA filters at $T_A=T_B=0.1$ have transmittivities $g_A \simeq 0.998$ and $g_B \simeq 0.0019$ where $g_A>0.5$ corresponds to NA on Alice's side and $g_B<0.5$ corresponds to NLA on Bob's side. Although these are labelled as different operations, their action on the single-rail basis is complementary and, up to an overall heralding-dependent scale, can be written as
\begin{equation}
M_A^{(\mathrm{NLA/NA})}\propto \mathrm{diag}\!\left(\sqrt{1-g_A},\sqrt{g_A}\right),
\end{equation}
\begin{equation}
M_B^{(\mathrm{NLA/NA})}\propto \mathrm{diag}\!\left(\sqrt{g_B},\sqrt{1-g_B}\right).
\end{equation}
Hence the corresponding POVM elements are
\begin{equation}
E_A^{(\mathrm{NLA/NA})}\propto \mathrm{diag}(1-g_A,g_A),
\end{equation}
\begin{equation}
E_B^{(\mathrm{NLA/NA})}\propto \mathrm{diag}(g_B,1-g_B),
\end{equation}
which, for the above transmittivities, are both approximately of the form $\mathrm{diag}(10^{-3},1)$. Therefore, despite the NA/NLA labels being opposite, both local filters perform the same physical task: they suppress the vacuum component $\ket{0}$ while preserving the single-photon component $\ket{1}$. This is consistent with the structure of the lossy shared state in Eq.~\eqref{eq:lossy_bell_state}, where loss introduces a vacuum admixture $|00\rangle\!\langle 00|$ in addition to the $\{|01\rangle,|10\rangle\}$ subspace. \ozlemREV{While this captures the main physical action of the NLA/NA filters, the constrained NLA/NA results remain slightly below the diagonal-filter in the low-transmission regime under the success-probability threshold used here. As discussed further in the following appendix, this difference is reduced when the NLA/NA success probability constraint is relaxed.}

\section{\label{appendix:optimisation_merit_psip}Optimisation Using Bell-State Fidelity}
\begin{figure}[t]
\hspace*{-0.7cm}
\includegraphics[scale=0.63]{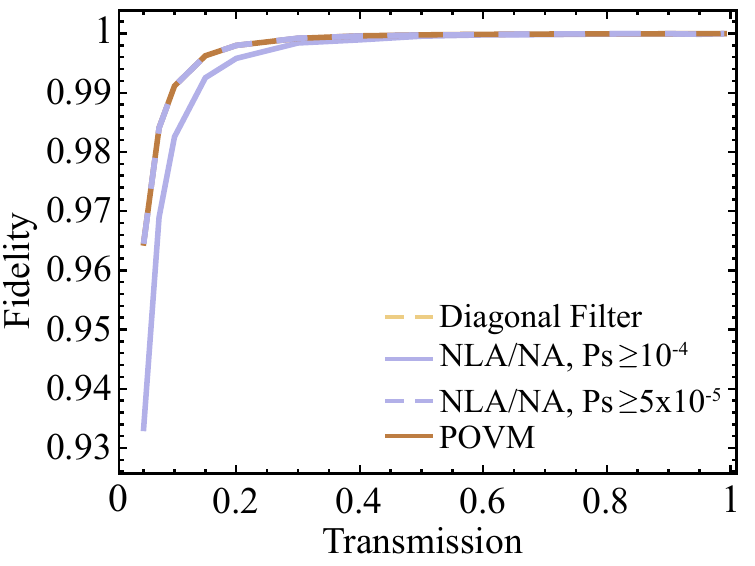}
\caption{\label{fig:figure_appendix2}Optimised conditional Bell-state fidelity $F_{\Psi^+}$ of the shared state between Alice and Bob after local filtering, shown as a function of the symmetric transmission $T_A = T_B$. We compare three filtering strategies: general POVMs, general diagonal filters, and NLA/NA filters, all optimised under the same success-probability constraint, $P_{\mathrm{succ}} \ge 10^{-4}$. The general diagonal filters closely track the optimised POVMs, and both show a modest advantage over the constrained NLA/NA filters at low transmission. The dashed blue curve shows the NLA/NA results when the success-probability threshold is relaxed to $P_{\mathrm{succ}} \ge 5 \times 10^{-5}$; in this case, the NLA/NA performance becomes comparable to the POVM and diagonal-filter curves in the low-transmission regime. At higher transmission, all approaches become nearly identical.}
\end{figure}
In this section, we benchmark local filtering strategies using a figure of merit defined directly on the shared state between Alice and Bob prior to teleportation. For each transmission setting, we optimise the fidelity of the post-filter shared state with the maximally entangled Bell state $\ket{\psi^+}$,
\begin{equation}
    \label{eq:fidelity_psip}
    F_{\psi^+}=\bra{\psi^+}\Tilde{\rho}_{AB}\ket{\psi^+},
\end{equation}
where $\Tilde{\rho}_{AB}$ is the renormalised shared state after successful local filtering on Alice and Bob. In both cases, the success probability is $P_\text{succ}\geq 10^{-4}$.

\ozlemREV{Figure~\ref{fig:figure_appendix2} shows the optimised $F_{\psi^+}$ results for the symmetric-loss case $T_A=T_B$ for different filtering strategies including the optimisation over general local diagonal filters. In the low-loss regime, the resulting diagonal-filter curves closely reproduce the full POVM results. By contrast, the NLA/NA family falls slightly below both the POVM and diagonal-filter optima in this regime when all three optimisations are performed under the same success-probability threshold, $P_{\mathrm{succ}} \ge 10^{-4}$. For comparison, we also repeated the NLA/NA optimisation with a relaxed threshold, $P_{\mathrm{succ}} \ge 5 \times 10^{-5}$. In this case, the NLA/NA curve becomes comparable to, and closely follows, the POVM and diagonal-filter curves in the low-transmission regime. This suggests that the previously observed gap is influenced by both the more restrictive structure of the constrained NLA/NA family and the success-probability threshold imposed on its optimisation. Relaxing the NLA/NA threshold substantially reduces this gap.}

\bibliography{apssamp}

\bibliographystyle{naturemag}

%
\end{document}